\documentclass[12pt,twoside,english,refpage,intoc,bibliography=totoc,index=totoc,BCOR7.5mm,captions=tableheading]{article}
\usepackage{lmodern}

\usepackage[T1]{fontenc}
\usepackage{textcomp}
\usepackage[latin9]{inputenc}
\usepackage{color}
\usepackage{babel}
\usepackage{mathrsfs}
\usepackage{amsmath}
\usepackage{geometry}
\usepackage[bookmarks=true,bookmarksnumbered=true,bookmarksopen=false,
 breaklinks=false,pdfborder={0 0 1},backref=false,colorlinks=true]
 {hyperref}
\hypersetup{pdftitle={The LyX User's Guide},
 pdfauthor={LyX Team},
 pdfsubject={LyX},
 pdfkeywords={LyX},
 linkcolor=black, citecolor=black, urlcolor=blue, filecolor=blue, pdfpagelayout=OneColumn, pdfnewwindow=true, pdfstartview=XYZ, plainpages=false}

\makeatletter
\numberwithin{equation}{section}
\newcommand{\lyxaddress}[1]{
	\par {\raggedright #1
	\vspace{1.4em}
	\noindent\par}
}

\usepackage{babel}
\usepackage[figure]{hypcap}

\usepackage{fancyhdr}

\@ifundefined{extrarowheight}{\usepackage{array}}{}
\usepackage{braket}
\makeatother

\begin{document}
\title{The double copy and perturbative teleparallel gravity}
\author{Georg Stettinger}
\maketitle

\lyxaddress{\begin{center}
\textcolor{black}{{International Center of Theoretical Sciences
(ICTS), 560089 Bengaluru, India} }
\par\end{center}}
\begin{abstract}
In this paper we analyze the double copy between gluon and graviton
amplitudes within the context of teleparallel gravity. We first set
up an effective Lagrangian based on minimal assumptions which reproduces
the on-shell amplitudes stemming from the double-copy procedure. By
straightforwardly computing the three- and four-point amplitude, this
double-copy Lagrangian can be constrained up to a few real parameters
when expanding to quartic order. For a specific choice of parameters,
we find an exact equivalence with perturbative teleparallel gravity.
This suggests that the teleparallel formulation might be well suited
to make the double copy more transparent from a Lagrangian viewpoint. 
\end{abstract}
\tableofcontents{}

\section{Introduction}

The \emph{double copy }is one of the most remarkable and fruitful
developments in the field of scattering amplitudes within the last
twenty years. In its simplest version, it states that gravity amplitudes
are related to gluon amplitudes in a very basic way, namely by replacing
the color factors by additional numerator factors, often summarized
by the slogan 
\[
\text{gravity}=\left(\text{Yang-Mills}\right)^{2}.
\]
See the original work \cite{Bern2010,Bern2008,Bern2010a} as well
as the excellent review \cite{Bern2024} and citations therein for
reference. Historically, the discovery goes back to the famous \emph{Kawai-Lewellen-Tye
(KLT) relations }\cite{Kawai1986} between open and closed string
amplitudes. Taking the low-energy limit gives rise to the above mentioned
relations for the massless sector. This is however not the full story
yet, meanwhile a whole web of theories connected by double-copy-type
relations has been discovered, including supergravities, non-linear
sigma models, Chern-Simons-theories and more, see \cite{Bern2024}\@. 

One of the reasons why the double copy seems surprising at first sight
is the arguably different appearance of the respective Lagrangians.
Whileas the Yang-Mills action 
\begin{equation}
S_{YM}=-\frac{1}{4}\int d^{4}x\,\,F_{\mu\nu}^{a}F_{a}^{\mu\nu}\label{eq:Yang-Mills action}
\end{equation}
is quadratic in the Yang-Mills curvature $F_{\mu\nu}^{a}$, the Einstein-Hilbert
action 
\begin{equation}
S_{EH}=-\frac{1}{16\pi}\int d^{4}x\,\sqrt{-g}R
\end{equation}
is linear in the spacetime curvature. This makes a Lagrangian explanation
difficult and not particularly transparent, some efforts in this direction
include \cite{Bern2010a,Tolotti2013}. The probably best Lagrangian
understanding of the double copy was put forward in \cite{Borsten2021},
see also the follow-up papers \cite{Borsten2022,Borsten2023,Borsten2023a}.
Here, the authors use the formalism of BRST-BV quantization and homotopy
algebras to prove the double copy relations in a general way, starting
from the Lagrangians. While the results are impressive, the arguments
require a high amount of heavy mathematical machinery. 

In this paper we want to make an effort to make the Lagrangian perspective
to the double copy more transparent by following a different route,
namely by realizing that the conventional formulation of Einstein
gravity is not the unique one. Teleparallel gravity based on torsion,
or more precisely, the \emph{teleparallel equivalent of Einstein gravity
(TEGR) }is perturbatively equivalent to general relativity but takes
a rather different form. The action can be written as 
\begin{equation}
S=-\frac{1}{32\pi}\int d^{4}x\,e\,T_{\,\,\,\mu\nu}^{a}T_{\,\,\,\rho\sigma}^{b}G_{a\,\,\,\,\,\,\,b}^{\,\,\,\mu\nu\,\,\,\rho\sigma},\label{eq:TEGR action-1}
\end{equation}
where $T_{\,\,\,\mu\nu}^{a}$ is the torsion tensor, $e$ is the determinant
of the vielbein and $G_{a\,\,\,\,\,\,\,b}^{\,\,\,\mu\nu\,\,\,\rho\sigma}$
is the super-metric, to be defined later. The similarity to (\ref{eq:Yang-Mills action})
is now obvious, especially when considering the fact that the torsion
tensor is also antisymmetric in $\mu$ and $\nu$ and has a form reminiscent
of a ``field strength''. There are important differences however:
In Yang-Mills, interactions come in due to the non-commutativity of
the gauge algebra, in TEGR the interactions stem from the expansions
of the super-metric and the volume element. As a result, TEGR has
vertices of arbitrary order, in contrast to the Yang-Mills action
which terminates at quartic order. Those differences still make a
direct Lagrangian comparison challenging. 

Instead, in this work we want to do the following: Let us suppose
we know the double-copy procedure and we know that it results in a
sensible physical theory. How should its Lagrangian look like? It
is of course impossible to obtain a unique Lagrangian from the knowledge
of the on-shell amplitudes only. The Lagrangian in general contains
a lot more information connected to field redefinitions, gauge invariance
and more. However, with a minimal amount of assumptions, we will construct
an ``effective double-copy Lagrangian'' up to quartic order, giving
rise to the amplitudes that result from the double-copy procedure.
To be as generic as possible, we will not use arguments such as gauge
invariance or diffeomorphism invariance in the ansatz. This effective
Lagrangian will be compared to (\ref{eq:TEGR action-1}): We will
see that it is not only compatible with TEGR, but also reproduces
its non-trivial content to a high extent. Motivated by this we will
argue that the teleparallel form of Einstein gravity is more suitable
to analyze and understand the double copy than the traditional form. 

The paper is organized as follows: In section two we give a short
review of both the double copy and teleparallel gravity to set the
notation and make everything self-contained. In section three we will
set up the effective double-copy action and discuss the necessary
assumptions and their consequences. Section four is devoted to explore
TEGR perturbatively and calculate the kinetic term and the cubic and
quartic vertex. Finally, in section five we compute the three- and
four-point amplitude using the effective action and constrain its
parameters by matching with the double-copy expressions. A comparison
with teleparallel gravity is added, together with a discussion of
what to expect from higher-order amplitudes. Section six just summarizes
the results and mentions some possible further directions.

\section{Review}

\subsection{Color-kinematics duality and the double copy}

We start with a short review of color-kinematics duality in Yang-Mills
theory. The action is given by 
\begin{equation}
S_{YM}=-\frac{1}{4}\int d^{4}x\,\,F_{\mu\nu}^{a}F_{a}^{\mu\nu}\label{eq:Yang Mills action}
\end{equation}
with the non-abelian field strength 
\begin{equation}
F_{\mu\nu}^{a}=\partial_{\mu}A_{\nu}^{a}-\partial_{\nu}A_{\mu}^{a}+gf_{\,\,\,bc}^{a}A_{\mu}^{b}A_{\nu}^{c}.
\end{equation}
Here, $A_{\mu}^{a}$ is the non-abelian gauge field in the adjoint
representation, $f^{abc}$ are the structure constants of the gauge
Lie algebra and $g$ is the just the Yang-Mills coupling constant.
Given any $m$-point tree-level on-shell amplitude, it can be expanded
as 
\begin{equation}
\mathcal{A}_{m}=-ig^{m-2}\sum_{i}\frac{c_{i}n_{i}}{D_{i}}\label{eq: amplitude decomposition}
\end{equation}
where the sum runs over all (tree-level) Feynman diagrams. The \emph{color
factors} $c_{i}$ are tensors in the adjoint of the Lie algebra composed
of the fundamental color tensors $f^{abc},$$t_{IJ}^{a}$ and possibly
also higher rank tensors. The \emph{numerators }$n_{i}$ are polynomials
in momenta, polarization vectors and spinors that also carry color
indices, dual to the $c_{i}$.\footnote{In this notation the color indices are suppressed to avoid cluttering
the formulas.} The \emph{denominators }$D_{i}$ come from the propagators involved
and are therefore products of terms of the form $p^{2}-m_{j}^{2}$
where the $m_{j}$ are the masses of the particles involved. While
it is easy to see from the structure of Feynman diagrams that such
a decomposition must exist, it is by no means unique. Consider a ``generalized
gauge transformation'' 
\begin{equation}
n_{i}\rightarrow n_{i}+\Delta_{i},\label{eq:generalized gauge}
\end{equation}
then the amplitude will remain invariant as long as the $\Delta_{i}$
fulfill 
\begin{equation}
\sum_{i}\frac{c_{i}\Delta_{i}}{D_{i}}=0.
\end{equation}

The color factors $c_{i}$ that arise are in general not independent,
but constrained through Lie algebra relations. It is instructive to
analyze that using the example of the tree-level on-shell 4-gluon
amplitude, which will play a significant role later on. We can decompose
it according to (\ref{eq: amplitude decomposition}) as 
\begin{equation}
\mathcal{A}_{4}=-ig^{2}\left(\frac{c_{s}n_{s}}{s}+\frac{c_{t}n_{t}}{t}+\frac{c_{u}n_{u}}{u}\right).
\end{equation}
Here, the denominators are just the standard Mandelstam variables
given by $s=\left(p_{1}+p_{2}\right)^{2},$$t=\left(p_{2}+p_{3}\right)^{2}$
and $u=\left(p_{1}+p_{3}\right)^{2}$. At this point one might ask,
what happened to the quartic interaction? In fact, it got absorbed
within the numerators as we will see below. The $c_{i}$ are given
by 
\begin{equation}
c_{s}=-2f^{a_{1}a_{2}b}f^{ba_{3}a_{4}},\,\,\,\,c_{t}=-2f^{a_{2}a_{3}b}f^{ba_{1}a_{4}},\,\,\,\,c_{u}=-2f^{a_{3}a_{1}b}f^{ba_{2}a_{4}}
\end{equation}
and obey 
\begin{equation}
c_{s}+c_{t}+c_{u}=0,\label{eq:color identity}
\end{equation}
which is just the Jacobi identity for the structure constants. Furthermore,
\begin{align}
n_{s}= & -\frac{1}{2}\Big(\left[\left(\epsilon_{1}^{a_{1}}\cdot\epsilon_{2}^{a_{2}}\right)p_{1}^{\mu}+2\left(\epsilon_{1}^{a_{1}}\cdot p_{2}\right)\epsilon_{2}^{a_{2}\mu}-\left(\epsilon_{2}^{a_{2}}\cdot\epsilon_{1}^{a_{1}}\right)p_{2}^{\mu}-2\left(\epsilon_{2}^{a_{2}}\cdot p_{1}\right)\epsilon_{1}^{a_{1}\mu}\right]\nonumber \\
 & \left[\left(\epsilon_{3}^{a_{3}}\cdot\epsilon_{4}^{a_{4}}\right)p_{3\mu}+2\left(\epsilon_{3}^{a_{3}}\cdot p_{4}\right)\epsilon_{4\mu}^{a_{4}}-\left(\epsilon_{4}^{a_{4}}\cdot\epsilon_{3}^{a_{3}}\right)p_{4\mu}-2\left(\epsilon_{4}^{a_{4}}\cdot p_{3}\right)\epsilon_{3\mu}^{a_{3}}\right]\nonumber \\
 & +s\left(\left(\epsilon_{1}^{a_{1}}\cdot\epsilon_{3}^{a_{3}}\right)\left(\epsilon_{2}^{a_{2}}\cdot\epsilon_{4}^{a_{4}}\right)-\left(\epsilon_{1}^{a_{1}}\cdot\epsilon_{4}^{a_{4}}\right)\left(\epsilon_{2}^{a_{2}}\cdot\epsilon_{3}^{a_{3}}\right)\right)\Big)\label{eq:n_s}
\end{align}
and $n_{t}$ and $n_{u}$ can be obtained by label permutations. The
polarization vectors $\epsilon_{i}^{a_{i}\mu}$ obey the on-shell
conditions $\epsilon_{i}\cdot\epsilon_{i}=0,$ $\epsilon_{i}\cdot p_{i}=0$.
The last line is coming from the quartic diagram which was omitted
in the decomposition: The prefactor $s$ just cancels the denominator
and what remains is the precise contribution that is needed. With
this trick one can recast any tree-level gluon amplitude in a purely
``cubic'' form with no elementary quartic vertex needed. 

The main statement of color-kinematics duality is now that not only
(\ref{eq:color identity}) holds, but also 
\begin{equation}
n_{s}+n_{t}+n_{u}=0,\label{eq:kin Jacobi}
\end{equation}
as can be verified directly. Similar relations are true for any order
as long as one is allowed to redefine the numerators as in (\ref{eq:generalized gauge}).
Precisely, what holds is the following: 

\emph{Every tree-level $n$-point gluon amplitude can be decomposed
as in (\ref{eq: amplitude decomposition}) such that the numerators
$n_{i}$ obey the same relations as the color factors $c_{i},$} \emph{which
follow from the Lie algebra structure.}

In fact, CK-duality holds in a much more general context, involving
loop diagrams and quark amplitudes, etc. For our purposes though,
the above statement is all we need. While this is already interesting
and highly non-trivial from a Lagrangian viewpoint, it also has a
profound consequence: It implies that one can get gauge invariant
objects by exchanging the color factors for an additional set of numerators,
schematically
\begin{equation}
c_{i}\rightarrow n_{i}.
\end{equation}
Consider the expression 
\begin{equation}
\mathcal{M}_{4}=-i\left(\frac{\kappa}{2}\right)^{2}\left(\frac{n_{s}^{2}}{s}+\frac{n_{t}^{2}}{t}+\frac{n_{u}^{2}}{u}\right),\label{eq:4-point grav amp}
\end{equation}
where the color indices are just omitted. This is nothing but the
tree-level 4-graviton amplitude in pure Einstein gravity up to an
overall normalization\footnote{The coupling needs to be adjusted, $g\rightarrow\frac{\kappa}{2}$
since the dimensions change.}. The external on-shell gravitons arise as tensor products of two
on-shell polarization vectors. This procedure of obtaining graviton
amplitudes from Yang-Mills amplitudes is called the \emph{double copy
}and has been shown to extend to all orders at tree-level. The proofs
are inductive and based on the BCFW recursion relations as well as
on unitarity methods. 

\subsection{Teleparallel gravity}

Teleparallel gravity is an alternative formulation of gravity which
is perturbatively equivalent to Einstein gravity. The main difference
is that it is based on torsion rather than curvature and therefore
structurewise closer to Yang-Mills theories, as we will see. The basic
field variable is the \emph{vielbein }$e_{\,\,\,\mu}^{a}$, which
relates the tangent space of spacetime to an abstract flat space,
called the fibre. It obeys the usual relations 
\begin{equation}
\eta_{ab}e_{\,\,\,\mu}^{a}e_{\,\,\,\nu}^{b}=g_{\mu\nu},\,\,\,\,\,\,\,\,\,\,\,\,\,\,\,\,\,\,\,\,g^{\mu\nu}e_{\,\,\,\mu}^{a}e_{\,\,\,\nu}^{b}=\eta^{ab},\,\,\,\,\,\,\,\,\,\,\,\,\,\,\,\,\,\,\,\,e_{\,\,\,\mu}^{a}e_{b}^{\,\,\,\mu}=\delta_{b}^{a},\,\,\,\,\,\,\,\,\,\,\,\,\,\,\,\,\,\,e_{\,\,\,\mu}^{a}e_{a}^{\,\,\,\nu}=\delta_{\mu}^{\nu}
\end{equation}
where $e_{a}^{\,\,\,\mu}$ is the inverse vielbein. Covariant derivatives
of vectors and one-forms living in the fibre are calculated using
the \emph{spin-connection }$\omega_{\,\,\,b\mu}^{a}$ as 
\begin{equation}
\mathcal{D}_{\mu}v^{a}=\partial_{\mu}v^{a}+\omega_{\,\,\,b\mu}^{a}v^{b},\,\,\,\,\,\,\,\,\,\,\,\,\,\,\,\,\,\mathcal{D}_{\mu}w_{a}=\partial_{\mu}w_{a}-\omega_{\,\,\,a\mu}^{b}w_{b}.
\end{equation}
Conversely, for spacetime vectors and one-forms the covariant derivative
reads 
\begin{equation}
\nabla_{\mu}v^{\nu}=\partial_{\mu}v^{\nu}+\Gamma_{\rho\mu}^{\nu}v^{\rho},\,\,\,\,\,\,\,\,\,\,\,\,\,\,\,\,\,\nabla_{\mu}w_{\nu}=\partial_{\mu}w_{\nu}-\Gamma_{\nu\mu}^{\rho}w_{\rho},
\end{equation}
using the general linear connection $\Gamma_{\mu\nu}^{\rho}.$ By
assumption, the two connections are related by the \emph{vielbein
postulate, }which simply states that the total covariant derivative
of $e_{\,\,\,\mu}^{a}$ vanishes: 
\begin{equation}
\partial_{\mu}e_{\,\,\,v}^{a}+\omega_{\,\,\,b\mu}^{a}e_{\,\,\,\nu}^{b}-\Gamma_{\nu\mu}^{\rho}e_{\,\,\,\rho}^{a}=0.
\end{equation}
This allows to express $\Gamma_{\mu\nu}^{\rho}$ in terms of $\omega_{\,\,\,b\mu}^{a}$
and vice versa, as long as the vielbein is fixed:
\begin{equation}
\Gamma_{\nu\mu}^{\rho}=e_{a}^{\,\,\,\rho}\partial_{\mu}e_{\,\,\,v}^{a}+e_{a}^{\,\,\,\rho}\omega_{\,\,\,b\mu}^{a}e_{\,\,\,\nu}^{b},
\end{equation}
\begin{equation}
\omega_{\,\,\,b\mu}^{a}=-e_{b}^{\,\,\,\nu}\partial_{\mu}e_{\,\,\,v}^{a}+e_{b}^{\,\,\,\nu}\Gamma_{\nu\mu}^{\rho}e_{\,\,\,\rho}^{a}.
\end{equation}
The curvature and torsion tensor can be calculated from the vielbein
and spin connection as 
\begin{equation}
R_{\,\,\,b\mu\nu}^{a}=\partial_{\mu}\omega_{\,\,\,b\nu}^{a}-\partial_{\nu}\omega_{\,\,\,b\mu}^{a}+\omega_{\,\,\,c\mu}^{a}\omega_{\,\,\,b\nu}^{c}-\omega_{\,\,\,c\nu}^{a}\omega_{\,\,\,b\mu}^{c}
\end{equation}
\begin{equation}
T_{\,\,\,\mu\nu}^{a}=\partial_{\mu}e_{\,\,\,\nu}^{a}-\partial_{\nu}e_{\,\,\,\mu}^{a}+\omega_{\,\,\,b\mu}^{a}e_{\,\,\,\nu}^{b}-\omega_{\,\,\,b\nu}^{a}e_{\,\,\,\mu}^{b},
\end{equation}
respectively. Their corresponding pure spacetime expressions read
\begin{equation}
R_{\,\,\,\sigma\mu\nu}^{\rho}=e_{a}^{\,\,\,\rho}e_{\,\,\,\sigma}^{b}R_{\,\,\,b\mu\nu}^{a}=\partial_{\mu}\Gamma_{\sigma\nu}^{\rho}-\partial_{\nu}\Gamma_{\sigma\mu}^{\rho}+\Gamma_{\lambda\mu}^{\rho}\Gamma_{\sigma\nu}^{\lambda}-\Gamma_{\lambda\nu}^{\rho}\Gamma_{\sigma\mu}^{\lambda},\label{eq:spacetime curvature}
\end{equation}
\begin{equation}
T_{\,\,\,\mu\nu}^{\rho}=e_{a}^{\,\,\,\rho}T_{\,\,\,\mu\nu}^{a}=\Gamma_{\nu\mu}^{\rho}-\Gamma_{\mu\nu}^{\rho}.
\end{equation}
In ordinary general relativity we demand that\footnote{Quantities associated to standard Einstein gravity, i. e. based on
the Levi-Civita connection, will be denoted with a small circle on
top.} 
\begin{equation}
T_{\,\,\,\mu\nu}^{\rho}\equiv\stackrel{\circ}{T^{\rho}}_{\,\,\,\mu\nu}=0,
\end{equation}
which together with metric compatibility leads to a unique solution
for the general connection called \emph{Levi-Civita-connection:}

\begin{equation}
\overset{\circ}{\Gamma}_{\nu\mu}^{\rho}=\frac{1}{2}g^{\rho\sigma}\left(\partial_{\nu}g_{\mu\sigma}+\partial_{\mu}g_{\nu\sigma}-\partial_{\sigma}g_{\mu\nu}\right).
\end{equation}
The curvature $\overset{\circ}{R}_{\,\,\,\sigma\mu\nu}^{\rho}$ computed
from $\overset{\circ}{\Gamma}_{\nu\mu}^{\rho}$ via equation (\ref{eq:spacetime curvature})
gives rise to the \emph{Einstein-Hilbert action }

\begin{equation}
S_{EH}=-\frac{1}{16\pi}\int d^{4}x\,\sqrt{-g}\overset{\circ}{R}.
\end{equation}
where $\overset{\circ}{R}=g^{\sigma\nu}\overset{\circ}{R}_{\,\,\,\sigma\mu\nu}^{\mu}$
is the \emph{Ricci scalar.}

In teleparallel gravity however, we do not restrict $T_{\,\,\,\mu\nu}^{a}$
but instead demand 

\begin{equation}
R_{\,\,\,b\mu\nu}^{a}=0
\end{equation}
identically, which immediately implies $R_{\,\,\,\sigma\mu\nu}^{\rho}=0.$
This means that, at least in a contractible neighbourhood, the spin
connection is ``pure gauge'' in the sense that it is equal to 
\begin{equation}
\omega_{\,\,\,b\mu}^{a}=\Lambda_{\,\,\,c}^{a}\partial_{\mu}\Lambda_{b}^{\,\,\,c}
\end{equation}
for some local Lorentz transformation $\Lambda_{\,\,\,b}^{a}\left(x^{\mu}\right)$.
One can now show by a direct calculation using the formulas above
(e.g. \cite{Aldrovandi2013,Arcos2004}) that the Ricci scalar\footnote{Quantities associated to standard Einstein gravity, i. e. based on
the Levi-Civita connection, will be denoted with a small circle on
top.} $\overset{\circ}{R}$ can be expressed in terms of the torsion tensor
as 
\begin{equation}
\overset{\circ}{R}=\frac{1}{4}T_{\,\,\,\mu\nu}^{a}T_{a}^{\,\,\,\mu\nu}+\frac{1}{2}T_{\,\,\,\mu\nu}^{a}T_{\,\,\,\,\,\,\,a}^{\nu\mu}-T_{\,\,\,\mu a}^{a}T_{\,\,\,\,\,\,\,b}^{b\mu}+\frac{2}{e}\partial_{\mu}\left(eT_{\,\,\,\,\,\,\,a}^{a\mu}\right)
\end{equation}
with $e=\text{det }e_{\,\,\,\mu}^{a}.$ This allows us to write the
Einstein-Hilbert action up to a boundary term as 
\begin{equation}
S_{EH}=-\frac{1}{32\pi}\int d^{4}x\,e\,\left(\frac{1}{2}T_{\,\,\,\mu\nu}^{a}T_{a}^{\,\,\,\mu\nu}+T_{\,\,\,\mu\nu}^{a}T_{\,\,\,\,\,\,\,a}^{\nu\mu}-2T_{\,\,\,\mu a}^{a}T_{\,\,\,\,\,\,\,b}^{b\mu}\right).
\end{equation}
For later purposes we will rewrite this as 
\begin{equation}
S=-\frac{1}{32\pi}\int d^{4}x\,e\,T_{\,\,\,\mu\nu}^{a}T_{\,\,\,\rho\sigma}^{b}G_{a\,\,\,\,\,\,\,b}^{\,\,\,\mu\nu\,\,\,\rho\sigma}\label{eq:TEGR action}
\end{equation}
where we introduced the ``super-metric'' 
\begin{equation}
G_{a\,\,\,\,\,\,\,b}^{\,\,\,\mu\nu\,\,\,\rho\sigma}=\frac{1}{2}\eta_{ab}g^{\mu\rho}g^{\nu\sigma}+e_{a}^{\,\,\,\sigma}e_{b}^{\,\,\,\nu}g^{\mu\rho}-2e_{a}^{\,\,\,\nu}e_{b}^{\,\,\,\sigma}g^{\mu\rho}.\label{eq:super metric}
\end{equation}
Varying the action with respect to $e_{\,\,\,\mu}^{a}$ and $\omega_{\,\,\,b\mu}^{a}$,
we notice that the equation of motion for the spin connection is identically
zero, such that the field can be considered pure gauge. There are
some caveats however related to global properties like instantons
for instance, see \cite{Krssak2025}. Still, since we are working
purely perturbatively here, we can safely set $\omega_{\,\,\,b\mu}^{a}$
to zero identically. This is called the \emph{teleparallel gauge,}
which indicates that the vielbeins are considered parallel over a
long distance. Now the torsion just takes the simple form 
\begin{equation}
T_{\,\,\,\mu\nu}^{a}=\partial_{\mu}e_{\,\,\,\nu}^{a}-\partial_{\nu}e_{\,\,\,\mu}^{a}.\label{eq:torsion def}
\end{equation}
Equation (\ref{eq:torsion def}) has the form of an abelian field
strength and indeed, there exists an interpretation of TEGR as a gauge
theory of the translation group \cite{Aldrovandi2013,Arcos2004}.
However, there has been raised criticism about its mathematical validity
and some very ``non-standard'' features compared to other gauge
theories, see \cite{LeDelliou2020,Brezina2026a,Huguet2021,Brezina2026}.
Nevertheless, thinking of $T_{\,\,\,\mu\nu}^{a}$ as some sort of
``field strength'' supports the comparison between the actions (\ref{eq:TEGR action})
and (\ref{eq:Yang Mills action}). Their obvious similarity is the
main motivation for this work.

\section{An effective double-copy Lagrangian}

The similarity between the Yang-Mills action and the teleparallel
gravity action suggests that one might be able to explain the double-copy
relations in a straightforward manner, directly from the Lagrangians.
This appears to be difficult though since there are still major differences
between the two actions:
\begin{itemize}
\item In (\ref{eq:Yang Mills action}), interactions are generated through
the commutator terms of the gauge field, leading to a cubic and a
quartic interaction vertex. In (\ref{eq:TEGR action}) the interactions
stem from the non-trivial, field-dependent super-metric (\ref{eq:super metric})
and the volume element, which lead to infinitely many vertices.
\item The kinetic part of (\ref{eq:Yang Mills action}) has one term whileas
that of (\ref{eq:TEGR action}) has three terms.
\item The volume element measure factor of $e$ appearing in (\ref{eq:TEGR action})
is missing in (\ref{eq:Yang Mills action}).
\end{itemize}
For those reasons it is quite difficult to find a direct derivation
of the double copy, based on a field redefinition for instance. Instead
we want to ask a different question: If we construct a Lagrangian
from the bottom up such that the double-copy relations in the form
of (\ref{eq:4-point grav amp}) hold to all orders, can we recover
gravity in the teleparallel formulation? Or more precisely: To what
extent can we recover gravity and what additional freedom remains?
This question we will try to answer in the next two sections.

\subsection{External particles}

Although the nature of the external particles involved is well-known,
we review the arguments here and convince ourselves that everything
follows from the double-copy properties only. From (\ref{eq:4-point grav amp}),
we see that the external polarization tensors $\epsilon_{\mu}^{a}$
all appear twice in the expression with different indices. Moreover,
we are instructed to omit the color index, hence we conclude that
the external particles of the resulting double copy theory should
have the form 
\begin{equation}
f_{\mu\nu}=\epsilon_{\mu}\epsilon_{\nu}.
\end{equation}
Since the $\epsilon_{\mu}^{a}$ were all null and transverse, our
new fields $f_{\mu\nu}$ and their associated momenta inherit the
following properties:
\begin{equation}
f_{\mu\nu}=f_{\nu\mu},\,\,\,\,\,\,\,\,\,\,\,\,f_{\mu}^{\mu}=0,\,\,\,\,\,\,\,\,\,\,\,\,\,\,p^{\mu}f_{\mu\nu}=0,\,\,\,\,\,\,\,\,\,\,\,\,\,p^{\mu}p_{\mu}=0.
\end{equation}
Those conditions define the $f_{\mu\nu}$ to be on-shell massless
gravitons. The two degrees of freedom contained in $\epsilon_{\mu}$
give rise to the polarizations of the graviton.

\subsection{Kinetic term}

Let us now make an ansatz for the kinetic term of our double-copy
Lagrangian. To stay as close as possible to the spirit of Yang-Mills
theory, let us first define the generalized ``field strength'' 
\begin{equation}
G_{\mu\nu\rho}=\partial_{\mu}f_{\nu\rho}-\partial_{\nu}f_{\mu\rho}.
\end{equation}
Now the simplest and most obvious guess for the kinetic term would
be 
\begin{equation}
S_{kin}^{naive}\propto\int d^{4}x\,\,G_{\mu\nu\rho}G^{\mu\nu\rho}.\label{eq:S_kin naive}
\end{equation}
However, this leads to inconsistencies: We know from the previous
section that $S_{kin}$ must be equal to the free action for massless
gravitons. This is unique up to partial integration and was first
derived by Fierz and Pauli. Its most convenient expression is given
in terms of spin projection operators (see Appendix) and reads 
\begin{equation}
S_{kin}\propto f_{\mu\nu}\mathcal{O}^{\mu\nu\rho\sigma}f_{\rho\sigma}\,\,\,\,\,\,\,\,\text{with\,\,\,\,\,\,\,\,\,\,}\mathcal{O}^{\mu\nu\rho\sigma}=p^{2}\left(P^{2}-2P_{s}^{0}\right).\label{eq:graviton prop}
\end{equation}
It is invariant under the infinitesimal linear gauge transformation
\begin{equation}
f_{\mu\nu}\rightarrow f_{\mu\nu}+\partial_{\mu}a_{\nu}+\partial_{\nu}a_{\mu}\label{eq:gauge sym}
\end{equation}
for some gauge parameter $a_{\mu}$. In gravity, this redundancy is
simply induced by diffeomorphism invariance. Its existence is also
supported by the double copy: Consider the expression for the four-point
amplitude (\ref{eq:4-point grav amp}) and replace one of the external
legs by $p_{\mu}\epsilon_{\nu}+\epsilon_{\mu}p_{\nu}:$ One gets 
\begin{equation}
-i\left(\frac{n_{s}^{2}}{s}+\frac{n_{t}^{2}}{t}+\frac{n_{u}^{2}}{u}\right)\mid_{\epsilon_{4\mu}\epsilon_{4\nu}\rightarrow p_{4\mu}\epsilon_{4\nu}+\epsilon_{4\mu}p_{4\nu}}=-i\left(n_{s}+n_{t}+n_{u}\right)\alpha\left(\epsilon_{i},p_{i}\right)
\end{equation}
for some function $\alpha\left(\epsilon_{i},p_{i}\right).$ This vanishes
according to the kinematic Jacobi identity (\ref{eq:kin Jacobi}),
which indicates that a field of the form $p_{\mu}\epsilon_{\nu}+\epsilon_{\mu}p_{\nu}$
is pure gauge. However, those Ward identities which hold for on-shell
amplitudes, typically do not capture the full information of gauge
redundancy. We therefore do not want to impose any requirements apart
from the double-copy relations directly, especially we will not demand
the full interacting theory to be diffeomorphism invariant. The linear
gauge redundancy of the free theory might extend to some non-linear
redundancy at higher orders, but we will not make any assumptions
about that. 

To make use of the spin projectors for our ansatz of $S_{kin}$ we
first go to Fourier space and define 
\begin{equation}
f_{\mu\nu}\left(x\right)=\int\frac{d^{4}p}{\left(2\pi\right)^{4}}F_{\mu\nu}\left(p\right)e^{ip_{\rho}x^{\rho}}.
\end{equation}
The kinetic operator corresponding to $S_{kin}^{naive}$ can now be
read off to be 
\begin{equation}
S_{kin}^{naive}\propto\int d^{4}x\,\,F^{\mu\nu}F^{\rho\sigma}\mathcal{O}_{\mu\nu\rho\sigma}^{kin}\,\,\,\Rightarrow\,\,\,\mathcal{O}_{\mu\nu\rho\sigma}^{kin}=p^{2}\eta_{\mu\rho}\eta_{\nu\sigma}-\eta_{\mu\rho}p_{\nu}p_{\sigma}.
\end{equation}
This can be expanded in the spin projectors and yields according to
the decomposition (\ref{eq:general spin decompo}) 
\begin{equation}
\mathcal{O}^{kin}=AP^{2}+BP^{1}+CP_{s}^{0}+DP_{w}^{0}+EP_{sw}^{0}+EP_{ws}^{0},
\end{equation}
\begin{equation}
A=p^{2},\,\,\,\,\,\,\,B=\frac{1}{2}p^{2},\,\,\,\,\,\,\,\,\,C=p^{2}.
\end{equation}
This is not compatible with the Fierz-Pauli action (\ref{eq:graviton prop}),
especially the appearance of longitudinal terms proportional to $P^{1}$
is disturbing. 

Since the simplest guess (\ref{eq:S_kin naive}) does not work, we
have to look for a more general form. Actually, $G_{\mu\nu\rho}$
is not the only ``field-strength-type'' expression we can write
down: All indices present are spacetime indices, so they can be arbitrarily
permuted. This yields the additional terms 
\begin{equation}
\partial_{\rho}f_{\nu\mu}-\partial_{\nu}f_{\rho\mu},\,\,\,\,\,\,\,,\left(\partial_{\mu}f_{\sigma}^{\sigma}-\partial_{\sigma}f_{\mu}^{\sigma}\right)\eta_{\nu\rho},\,\,\,\,\,\,\,\partial_{\mu}f_{\rho\nu}-\partial_{\rho}f_{\mu\nu.}
\end{equation}
The last one however turns out to be redundant, so we can write our
generalized ansatz as 
\begin{align}
S_{kin}^ {} & \propto\frac{1}{2}\int d^{4}x\,\,\left(\partial_{\mu}f_{\nu\rho}-\partial_{v}f_{\mu\rho}+\alpha\left(\partial_{\rho}f_{\nu\mu}-\partial_{\nu}f_{\rho\mu}\right)+\beta\left(\partial_{\mu}f_{\sigma}^{\sigma}-\partial_{\sigma}f_{\mu}^{\sigma}\right)\eta_{\nu\rho}\right)\nonumber \\
 & \,\,\,\,\,\,\,\,\,\,\,\,\,\,\,\,\,\left(\partial^{\mu}f^{\nu\rho}-\partial^{\nu}f^{\mu\rho}+\gamma\left(\partial^{\rho}f^{\nu\mu}-\partial^{\nu}f^{\rho\mu}\right)+\delta\left(\partial^{\mu}f_{\sigma}^{\sigma}-\partial_{\sigma}f^{\sigma\mu}\right)\eta^{\nu\rho}\right).
\end{align}
for arbitrary coefficients $\alpha$, $\beta$, $\gamma$, $\delta$.
This may at first sight seem almost too general: One might be tempted
to choose a \emph{symmetric ansatz }where $\alpha=\gamma$ and $\beta=\delta$.
But as we will see, the symmetric ansatz would not lead to a satisfying
solution. It is straightforward to compute the kinetic operator $\mathcal{O}_{\mu\nu\rho\sigma}^{kin}$
in Fourier space in terms of the coefficients:
\begin{align}
\mathcal{O}_{\mu\nu\rho\sigma}^{kin}F^{\mu\nu}\left(p\right)F^{\rho\sigma}\left(-p\right) & =F^{\mu\nu}\left(p\right)F^{\rho\sigma}\left(-p\right)\Big[\frac{1}{2}p^{2}\left(\eta_{\mu\rho}\eta_{\nu\sigma}+\eta_{\mu\sigma}\eta_{\nu\rho}\right)\left(2+\alpha+\gamma+2\alpha\gamma\right)+p^{2}\eta_{\mu\nu}\eta_{\rho\sigma}\left(\beta+\delta+4\beta\delta\right)\nonumber \\
 & \,\,\,\,\,\,\,\,\,+\frac{1}{2}\left(p_{\mu}p_{\nu}\eta_{\rho\sigma}+p_{\rho}p_{\sigma}\eta_{\mu\nu}\right)\left(-2\beta-2\delta-8\beta\delta\right)\nonumber \\
 & \,\,\,\,\,\,\,\,\,\,+\frac{1}{4}\left(p_{\mu}p_{\rho}\eta_{\nu\sigma}+p_{\mu}p_{\sigma}\eta_{\nu\rho}+p_{\nu}p_{\sigma}\eta_{\mu\rho}+p_{\nu}p_{\rho}\eta_{\mu\sigma}\right)\left(-2-\alpha-\gamma-2\alpha\gamma+\beta\delta+4\beta\delta\right)\Big]
\end{align}
Expanding in the spin projectors as above yields 
\begin{align}
A & =p^{2}\left(2+\alpha+\gamma+2\alpha\gamma\right)\\
B & =p^{2}\left(1+\frac{\alpha}{2}+\frac{\gamma}{2}+\alpha\gamma+\frac{\beta}{2}+\frac{\delta}{2}+2\beta\delta\right)\\
C & =p^{2}\left(2+\alpha+\gamma+2\alpha\gamma+3\beta+3\delta+12\beta\delta\right)\\
D & =E=0.
\end{align}
We see that even for this very general ansatz, three coefficients
that we need to be zero, already vanish identically. Still there are
two conditions left: We need $B$ to vanish since it corresponds to
the unphysical longitudinal modes and we need the correct ratio between
$A$ and $C.$ Since we have four unknown constants, we expect some
freedom of choice: Let us first explore the above mentioned symmetric
ansatz and set $\alpha=\gamma$, $\beta=\delta$. The equation for
$B$ we need to solve now becomes 
\begin{equation}
1+\alpha+\alpha^{2}+\beta+2\beta^{2}=0.
\end{equation}
It is easy to see that there are no solutions within the real numbers,
so the symmetric ansatz has to be abandoned. 

A second natural ansatz and the one that we will adopt instead, is
to set $\gamma=\delta=0,$ such that one of the two factors in the
kinetic term is just the ``field strength'' $G_{\mu\nu\rho}$. This
also leads to a straightforward way to include interactions: We just
consider an action of the form 
\begin{equation}
S\propto\int d^{4}x\,\,G_{\mu\nu\rho}S^{\mu\nu\rho}\label{eq:proposed action}
\end{equation}
where $S^{\mu\nu\rho}$ is some tensor, antisymmetric in $\mu$ and
$\nu$, constructed out of $f_{\mu\nu}$ and its derivatives. The
$\mathcal{O}\left(f_{\mu\nu}\right)$-contribution gives rise to the
kinetic term.

The equations to be solved now become 
\begin{equation}
1+\frac{\alpha}{2}+\frac{\beta}{2}=0,\,\,\,\,\,\,\,\,\,,\frac{2+\alpha+3\beta}{2+\alpha}=-2.
\end{equation}
They are in fact dependent on each other and both solved by 
\begin{equation}
\beta=-\alpha-2,
\end{equation}
which yields 
\begin{equation}
A=p^{2}\left(\alpha+2\right),\,\,\,\,\,\,\,\,\,\,\,\,C=-p^{2}\left(2\alpha+4\right)
\end{equation}
and all others vanishing. Therefore the last remaining constant $\alpha$
only appears within an overall prefactor and cannot be determined
yet. With the choice $\alpha=2$ we get\footnote{From now on the trace $f_{\mu}^{\mu}$ shall be denoted as $f$.}
\begin{equation}
S^{\mu\nu\rho}=\frac{1}{2}\left(\partial^{\mu}f^{\nu\rho}-\partial^{\nu}f^{\mu\rho}\right)+\partial^{\rho}f^{\nu\mu}-\partial^{\nu}f^{\rho\mu}-2\left(\partial^{\mu}f-\partial_{\sigma}f^{\sigma\mu}\right)\eta^{\nu\rho}+\mathcal{O}\left(f_{\mu\nu}^{2}\right).\label{eq:excitation form linear}
\end{equation}
At this point one can already compare the ratios of the coefficients
to (\ref{eq:super metric}). In anticipation of the comparison to
perturbative TEGR, we will choose $\alpha=2$ as a normalization condition
on the action. The propagator can then per construction be chosen
as proportional to the Feynman gauge propagator\footnote{In principle, using the Feynman propagator requires the gauge invariance
(\ref{eq:gauge sym}) to hold to all orders. Since we do not assume
that, the use of this propagator should be seen as part of the prescription. } and reads 
\begin{equation}
\Pi_{GR}=\frac{1}{8p^{2}}\left(\eta_{\mu\rho}\eta_{\nu\sigma}+\eta_{\mu\sigma}\eta_{\nu\rho}-\eta_{\mu\nu}\eta_{\rho\sigma}\right).\label{eq:prop}
\end{equation}
One should notice that this choice does not correspond to the canonical
normalization, see section 5.

\subsection{Cubic vertex\label{subsec:Cubic-term}}

Given the proposed form of the action (\ref{eq:proposed action})
we will now make a general ansatz for the quadratic part of $S_{\mu\nu\rho}.$
As already emphasized, we want to leave the interacting part of the
theory completely general and do not impose any kind of gauge invariance. 

We know from (\ref{eq:4-point grav amp}) that it should contain one
derivative and of course two $f_{\mu\nu}$-fields. Taking into account
the antisymmetry in $\mu$ and $\nu$, it can be expanded in thirteen
terms: 
\begin{align}
S_{\mu\nu\rho}\mid_{qu}\,=\, & c_{1}\left(f\partial_{\mu}f\eta_{\nu\rho}-f\partial_{\nu}f\eta_{\mu\rho}\right)+c_{2}\left(f^{\sigma\lambda}\partial_{\mu}f_{\sigma\lambda}\eta_{\nu\rho}-f^{\sigma\lambda}\partial_{\nu}f_{\sigma\lambda}\eta_{\mu\rho}\right)+c_{3}\left(f\partial_{\mu}f_{\nu\rho}-f\partial_{\nu}f_{\mu\rho}\right)+\nonumber \\
 & c_{4}\left(f_{\nu\rho}\partial_{\mu}f-f_{\mu\rho}\partial_{\nu}f\right)+c_{5}\left(f_{\rho}^{\sigma}\partial_{\mu}f_{\nu\sigma}-f_{\rho}^{\sigma}\partial_{\nu}f_{\mu\sigma}\right)+c_{6}\left(f_{\nu}^{\sigma}\partial_{\mu}f_{\rho\sigma}-f_{\mu}^{\sigma}\partial_{\nu}f_{\rho\sigma}\right)+\nonumber \\
 & c_{7}\left(f_{\mu\sigma}\partial_{\rho}f_{\nu}^{\sigma}-f_{\nu\sigma}\partial_{\rho}f_{\mu}^{\sigma}\right)+c_{8}\left(f_{\mu\rho}\partial_{\sigma}f_{\nu}^{\sigma}-f_{\nu\rho}\partial_{\sigma}f_{\mu}^{\sigma}\right)+c_{9}\left(f_{\mu}^{\sigma}\partial_{\sigma}f_{\nu\rho}-f_{\nu}^{\sigma}\partial_{\sigma}f_{\mu\rho}\right)+\nonumber \\
 & c_{10}\left(f\partial_{\sigma}f_{\mu}^{\sigma}\eta_{\nu\rho}-f\partial_{\sigma}f_{\nu}^{\sigma}\eta_{\mu\rho}\right)+c_{11}\left(f_{\mu}^{\sigma}\partial_{\sigma}f\eta_{\nu\rho}-f_{\nu}^{\sigma}\partial_{\sigma}f\eta_{\mu\rho}\right)+\nonumber \\
 & c_{12}\left(f_{\mu}^{\sigma}\partial_{\lambda}f_{\sigma}^{\lambda}\eta_{\nu\rho}-f_{\nu}^{\sigma}\partial_{\lambda}f_{\sigma}^{\lambda}\eta_{\mu\rho}\right)+c_{13}\left(f_{\sigma}^{\lambda}\partial_{\lambda}f_{\mu}^{\sigma}\eta_{\nu\rho}-f_{\sigma}^{\lambda}\partial_{\lambda}f_{\nu}^{\sigma}\eta_{\mu\rho}\right).\label{eq:c_decomposition}
\end{align}
In a similar way one can analyze the full three-vertex $V_{3}$. This
time there are sixteen different possible index contractions of three
fields and two derivatives. The ansatz (\ref{eq:proposed action})
dictates that the derivatives have to act on different fields, hence
we get the general expression
\begin{align}
V_{3}=\, & d_{1}f\partial_{\mu}f^{\mu\nu}\partial_{\nu}f+d_{2}f\partial_{\mu}f^{\mu\nu}\partial_{\rho}f_{\nu}^{\rho}+d_{3}f\partial_{\mu}f\partial^{\mu}f+d_{4}f\partial_{\mu}f^{\nu\rho}\partial^{\mu}f_{\nu\rho}+\nonumber \\
 & d_{5}f\partial_{\mu}f^{\nu\rho}\partial_{\nu}f_{\rho}^{\mu}+d_{6}f_{\mu\nu}\partial^{\mu}f^{\nu\rho}\partial_{\rho}f+d_{7}f_{\mu\nu}\partial^{\mu}f^{\nu\rho}\partial_{\sigma}f_{\rho}^{\sigma}+d_{8}f_{\mu\nu}\partial^{\mu}f\partial^{\nu}f+\nonumber \\
 & d_{9}f_{\mu\nu}\partial^{\mu}f^{\rho\sigma}\partial^{\nu}f_{\rho\sigma}+d_{10}f_{\mu\nu}\partial^{\mu}f\partial_{\rho}f^{\nu\rho}+d_{11}f_{\mu\nu}\partial^{\mu}f_{\rho\sigma}\partial^{\rho}f^{\nu\sigma}+d_{12}f_{\mu\nu}\partial^{\rho}f^{\mu\nu}\partial_{\rho}f+\nonumber \\
 & d_{13}f_{\mu\nu}\partial_{\rho}f^{\mu\nu}\partial_{\sigma}f^{\rho\sigma}+d_{14}f_{\mu\nu}\partial_{\rho}f^{\mu\rho}\partial_{\sigma}f^{\nu\sigma}+d_{15}f_{\mu\nu}\partial_{\rho}f^{\mu\sigma}\partial^{\rho}f_{\sigma}^{\nu}+d_{16}f_{\mu\nu}\partial_{\rho}f^{\nu\sigma}\partial_{\sigma}f^{\mu\rho}.\label{eq:general 3-vertex}
\end{align}
 After contracting $S_{\mu\nu\rho}\mid_{qu}$ with $G^{\mu\nu\rho}$
we can calculate the $d_{i}$ in terms of the $c_{i}$ which leads
to 
\begin{align}
 & d_{1}=-2c_{1}+2c_{10},\,\,\,\,\,\,\,d_{2}=-2c_{10},\,\,\,\,\,\,\,\,\,\,\,\,\,\,\,\,\,\,\,d_{3}=2c_{1},\,\,\,\,\,\,\,\,\,\,\,\,\,\,\quad\,\,\,\,\,\,\,\,\,\,\,\,\,\,\,d_{4}=2c_{3},\nonumber \\
 & d_{5}=-2c_{3},\,\,\,\,\,\,\:\,\,\,\,\,\,\,\,\,\,\,\,\,d_{6}=-2c_{4}+2c_{13},\,\,\,\,\,\,\,\,d_{7}=2c_{8}-2c_{13},\,\,\,\,\,\,\,\,\,\,\,\,\,\,\,\,\,\,\,d_{8}=2c_{11},\nonumber \\
 & d_{9}=2c_{9},\,\,\,\,\,\,\,\,\,\,\,\,\,\,\,\,\,\,\,\,\,\,\,\,d_{10}=-2c_{11}+2c_{12},\,\,\,\,\,d_{11}=-2c_{6}+2c_{7}-2c_{9},\,\,\,\,d_{12}=2c_{2}+2c_{4},\nonumber \\
 & d_{13}=-2c_{2}-2c_{8},\,\,\,\,\,\,\,\,d_{14}=-2c_{12},\,\,\,\,\,\,\,\,\,\,\,\,\,\,\,\,\,\,d_{15}=2c_{5}+2c_{6},\,\,\,\,\,\,\,\,\,\,\,\,\,\,\,\,\,\,\,d_{16}=-2c_{5}-2c_{7}.\label{eq:c-d relation}
\end{align}
The factors of two obviously stem from the antisymmetrization. We
see that the relation between the $d_{i}$ and the c$_{i}$ is not
invertible: We have the linear relations 
\begin{align}
d_{1}+d_{2}+d_{3}=0,\,\,\,\,\,\,\,\,\,\,d_{4}+d_{5}=0,\,\,\,\,\,\,\,\,\,\,\,d_{6}+d_{7}+d_{12}+d_{13}=0,\nonumber \\
d_{8}+d_{10}+d_{14}=0,\,\,\,\,\,\,\,\,\,\,\,d_{9}+d_{11}+d_{15}+d_{16}=0.\label{eq:d_relations}
\end{align}
This means that the ansatz (\ref{eq:proposed action}) is already
restrictive and cannot reproduce the most general 3-vertex. For the
calculation of the amplitudes in the next section it is useful to
define a few subsets of terms of the general expression (\ref{eq:general 3-vertex}):
\begin{equation}
\mathscr{V}_{rel4}=\left\{ d_{4},d_{5},d_{6},d_{7},d_{9},d_{11},d_{12},d_{13},d_{15},d_{16}\right\} \subset\mathscr{V}_{trans}
\end{equation}
These terms contain at most one trace or divergence factor and will
be those which are relevant for the 4-point amplitude calculation.
Since the external states are transverse-traceless, only one leg of
the 3-vertex, namely the one connected to the propagator, can form
a trace or divergence to get a non-vanishing result.
\begin{equation}
\mathscr{V}_{rel3}=\left\{ d_{9},d_{11},d_{15},d_{16}\right\} \subset\mathscr{V}_{rel4}
\end{equation}
These terms are relevant for the 3-point amplitude and contain neither
a trace nor a divergence. 
\begin{equation}
\mathscr{V}_{ind}=\left\{ d_{2},d_{3},d_{4},d_{6},d_{9},d_{10},d_{11},d_{12},d_{13},d_{14},d_{16}\right\} 
\end{equation}
These terms form an independent set under the relations (\ref{eq:d_relations}).
The coefficients that we will actually use in the calculation of the
3-point amplitude (4-point amplitude) are therefore $\mathscr{V}_{rel3}\cap\mathscr{V}_{ind}$
$\left(\mathscr{V}_{rel4}\cap\mathscr{V}_{ind}\right)$. 

\subsection{Quartic vertex}

In a similar fashion we now want to construct the quartic interaction
of the proposed action. At this stage it is already meaningful to
implement some simplifications: Since the highest amplitude we will
compute is the 4-point amplitude, we know that all terms which contain
either a trace or a divergence, will be annihilated by the external
on-shell states. We can therefore omit them from the beginning and
start in the following way, in accordance to the previous section:
We first write down the most general cubic part of $S_{\mu\nu\rho}$
subject to the criteria:
\begin{enumerate}
\item It has to contain three $f_{\mu\nu}$- fields and one derivative.
\item It has to be antisymmetric in $\mu$ and $\nu$.
\item It must not contain any trace, divergence or $\eta_{\mu\nu}$ (that
would lead to a trace or divergence after contraction with $G_{\mu\nu\rho}$).
\end{enumerate}
The result is 
\begin{align}
S_{\mu\nu\rho}\mid_{cub}\,=\, & k_{1}\left(\partial_{\mu}f_{\nu\rho}f_{\sigma\lambda}f^{\sigma\lambda}-\partial_{\nu}f_{\mu\rho}f_{\sigma\lambda}f^{\sigma\lambda}\right)+k_{2}\left(\partial_{\mu}f_{\nu\sigma}f_{\rho\lambda}f^{\sigma\lambda}-\partial_{\nu}f_{\mu\sigma}f_{\rho\lambda}f^{\sigma\lambda}\right)\nonumber \\
+ & k_{3}\left(\partial_{\mu}f_{\rho\sigma}f_{\nu\lambda}f^{\sigma\lambda}-\partial_{\nu}f_{\rho\sigma}f_{\mu\lambda}f^{\sigma\lambda}\right)+k_{4}\left(\partial_{\rho}f_{\mu\sigma}f_{\nu\lambda}f^{\sigma\lambda}-\partial_{\rho}f_{\nu\sigma}f_{\mu\lambda}f^{\sigma\lambda}\right)\nonumber \\
+ & k_{5}\left(\partial_{\mu}f_{\sigma\lambda}f_{\nu\rho}f^{\sigma\lambda}-\partial_{\nu}f_{\sigma\lambda}f_{\mu\rho}f^{\sigma\lambda}\right)+k_{6}\left(\partial_{\mu}f_{\sigma\lambda}f_{\nu}^{\,\,\,\sigma}f_{\rho}^{\,\,\,\lambda}-\partial_{\nu}f_{\sigma\lambda}f_{\mu}^{\,\,\,\sigma}f_{\rho}^{\,\,\,\lambda}\right)\nonumber \\
+ & k_{7}\left(\partial_{\sigma}f_{\mu\rho}f_{\nu\lambda}f^{\sigma\lambda}-\partial_{\sigma}f_{\nu\rho}f_{\mu\lambda}f^{\sigma\lambda}\right)+k_{8}\left(\partial_{\sigma}f_{\mu\lambda}f_{\nu\rho}f^{\sigma\lambda}-\partial_{\sigma}f_{\nu\lambda}f_{\mu\rho}f^{\sigma\lambda}\right)\nonumber \\
+ & k_{9}\left(\partial_{\sigma}f_{\mu\lambda}f_{\rho}^{\,\,\,\lambda}f_{\nu}^{\,\,\,\sigma}-\partial_{\sigma}f_{\nu\lambda}f_{\rho}^{\,\,\,\lambda}f_{\nu}^{\,\,\,\sigma}\right)+k_{10}\left(\partial_{\sigma}f_{\mu\lambda}f_{\nu}^{\,\,\,\lambda}f_{\rho}^{\,\,\,\sigma}-\partial_{\sigma}f_{\nu\lambda}f_{\mu}^{\,\,\,\lambda}f_{\rho}^{\,\,\,\sigma}\right)\nonumber \\
+ & k_{11}\left(\partial_{\sigma}f_{\rho\lambda}f_{\mu}^{\,\,\,\lambda}f_{\nu}^{\,\,\,\sigma}-\partial_{\sigma}f_{\rho\lambda}f_{\nu}^{\,\,\,\lambda}f_{\mu}^{\,\,\,\sigma}\right)
\end{align}
On the other hand, the most general four-vertex with no traces and
no divergences, where the two derivatives act on different fields,
is given by 
\begin{align}
V_{4}^{on-shell}=\, & l_{1}\partial_{\mu}f_{\nu\rho}\partial^{\mu}f^{\nu\rho}f_{\sigma\lambda}f^{\sigma\lambda}+l_{2}\partial_{\mu}f_{\nu\rho}\partial^{\mu}f^{\nu\sigma}f_{\sigma\lambda}f^{\rho\lambda}+l_{3}\partial_{\mu}f_{\nu\rho}\partial^{\mu}f^{\sigma\lambda}f^{\nu\rho}f_{\sigma\lambda}\nonumber \\
+ & l_{4}\partial_{\mu}f_{\nu\rho}\partial^{\mu}f^{\sigma\lambda}f_{\,\,\,\sigma}^{\nu}f_{\,\,\,\lambda}^{\rho}+l_{5}\partial_{\mu}f_{\nu\rho}\partial^{\nu}f^{\mu\rho}f_{\sigma\lambda}f^{\sigma\lambda}+l_{6}\partial_{\mu}f_{\nu\rho}\partial^{\nu}f^{\mu\sigma}f^{\rho\lambda}f_{\sigma\lambda}\nonumber \\
+ & l_{7}\partial_{\mu}f_{\nu\rho}\partial^{\nu}f^{\rho\sigma}f_{\sigma\lambda}f^{\mu\lambda}+l_{8}\partial_{\mu}f_{\nu\rho}\partial^{\nu}f^{\sigma\lambda}f^{\mu\rho}f_{\sigma\lambda}+l_{9}\partial_{\mu}f_{\nu\rho}\partial^{\nu}f^{\sigma\lambda}f_{\,\,\,\sigma}^{\mu}f_{\,\,\,\lambda}^{\rho}\nonumber \\
+ & l_{10}\partial_{\mu}f_{\nu\rho}\partial^{\sigma}f^{\nu\rho}f^{\mu\lambda}f_{\sigma\lambda}+l_{11}\partial_{\mu}f_{\nu\rho}\partial^{\sigma}f^{\nu\lambda}f^{\mu\rho}f_{\sigma\lambda}+l_{12}\partial_{\mu}f_{\nu\rho}\partial^{\sigma}f^{\nu\lambda}f_{\,\,\,\sigma}^{\mu}f_{\,\,\,\lambda}^{\rho}\nonumber \\
+ & l_{13}\partial_{\mu}f_{\nu\rho}\partial^{\nu}f^{\sigma\lambda}f_{\,\,\,\lambda}^{\mu}f_{\,\,\,\sigma}^{\rho}.\label{eq:most gen 4vert}
\end{align}
Comparing the two expressions again leads to a few linear relations:
\begin{equation}
l_{1}+l_{5}=0,\,\,\,\,\,\,\,\,\,\,\,\,\,\,l_{2}+l_{6}+l_{7}+l_{10}=0,\,\,\,\,\,\,\,\,\,\,\,\,\,\,\,l_{3}+l_{8}+l_{11}=0,\,\,\,\,\,\,\,\,\,\,\,\,\,l_{4}+l_{9}+l_{12}+l_{13}=0.\label{eq:l-relations}
\end{equation}
A set of independent coefficients for the amplitude calculation is
given by 
\begin{equation}
\mathcal{V}_{ind}=\left\{ l_{5},l_{6},l_{7},l_{8},l_{9},l_{10},l_{11},l_{12},l_{13}\right\} .
\end{equation}

\section{Perturbative teleparallel gravity}

To be able to compare our results to TEGR, we should know what to
expect, hence we need to do perturbation theory and identify the graviton
interactions. Perturbative TEGR has not been discussed much in the
literature so far, see for instance \cite{Landry2023}, hence there
might be some new results in this section. We start with the following
decomposition of the vielbein: 
\begin{equation}
e_{\,\,\,\mu}^{a}=\delta_{\,\,\,\mu}^{a}+f_{\,\,\,\mu}^{a}.\label{eq:vielbein expansion}
\end{equation}
This is a non-trivial equation, since it implies that we are doing
perturbation theory around flat Minkowski background. Calculating
the metric we get 
\begin{equation}
g_{\mu\nu}=\eta_{ab}\left(\delta_{\,\,\,\mu}^{a}+f_{\,\,\,\mu}^{a}\right)\left(\delta_{\,\,\,\nu}^{b}+f_{\,\,\,\nu}^{b}\right)=\eta_{\mu\nu}+f_{\mu\nu}+f_{\nu\mu}+f_{\,\,\,\mu}^{a}f_{a\nu}=:\eta_{\mu\nu}+h_{\mu\nu}
\end{equation}
which implies that we can identify the standard graviton as 
\begin{equation}
h_{\mu\nu}=f_{\mu\nu}+f_{\nu\mu}+f_{\,\,\,\mu}^{a}f_{a\nu}.
\end{equation}
To first order in $f_{\,\,\,\mu}^{a}$ it is just two times the symmetrized
vielbein. This makes sense, since the antisymmetric part is associated
to the additional gauge freedom of local Lorentz symmetry and not
physical. We will therefore demand 
\begin{equation}
f_{\left[\mu\nu\right]}=0
\end{equation}
and work with the symmetric part only. Below we will see that the
antisymmetric part does not propagate and drops out of all vertices.
The inverse vielbein can be expanded as 
\begin{equation}
e_{a}^{\,\,\,\mu}=\delta_{a}^{\,\,\,\mu}-f_{a}^{\,\,\,\mu}+f_{a}^{\,\,\,\nu}f_{\nu}^{\,\,\,\mu}+\mathcal{O}\left(f^{3}\right).\label{eq:inverse vielbein}
\end{equation}
Note that $e_{a}^{\,\,\,\mu}$ gains corrections of \emph{all }orders
in $f_{a}^{\,\,\,\mu}$, which, as we will see, leads to the infinitely
many vertices of perturbative TEGR.

\subsection{Kinetic term}

We start from the action in the form (\ref{eq:TEGR action})\footnote{At this point, it is notationally convenient to express everything
in terms of spacetime indices only, using the background field $\delta_{\,\,\,\mu}^{a}$
which serves as an isomorphism between the tangent space and the fibre.}:
\begin{equation}
S=-\frac{1}{32\pi}\int d^{4}x\,e\,T_{\,\,\,\mu\nu}^{\lambda}T_{\,\,\,\rho\sigma}^{\kappa}G_{\lambda\,\,\,\,\,\,\,\kappa}^{\,\,\,\mu\nu\,\,\,\rho\sigma}.
\end{equation}
 The torsion tensor, expanded according to (\ref{eq:vielbein expansion}),
is already exact to first order in $f_{\,\,\,\mu}^{\lambda}$ and
reads 
\begin{equation}
T_{\,\,\,\mu\nu}^{\lambda}=\partial_{\mu}f_{\,\,\,\nu}^{\lambda}-\partial_{\nu}f_{\,\,\,\mu}^{\lambda}.
\end{equation}
This means, for the kinetic term we do not need to expand the volume
element and the super-metric. Instead, we just evaluate them on Minkowski
space and obtain
\begin{equation}
e\mid_{f^{0}}=1
\end{equation}
\begin{equation}
G_{\lambda\,\,\,\,\,\,\,\kappa}^{\,\,\,\mu\nu\,\,\,\rho\sigma}\mid_{f^{0}}=\frac{1}{2}\eta_{\lambda\kappa}\eta^{\mu\rho}\eta^{\nu\sigma}+\delta_{\lambda}^{\,\,\,\sigma}\delta_{\kappa}^{\,\,\,\nu}\eta^{\mu\rho}-2\delta_{\lambda}^{\,\,\,\nu}\delta_{\kappa}^{\,\,\,\sigma}\eta^{\mu\rho}\label{eq:super metric flat}
\end{equation}
 Multiplying everything out yields 
\begin{equation}
S_{kin}=-\frac{1}{16\pi}\int d^{4}x\,\,\left[\partial_{\mu}f_{\,\,\,\nu}^{\lambda}\partial^{\mu}f_{\lambda}^{\,\,\,\nu}-\partial_{\nu}f_{\,\,\,\mu}^{\lambda}\partial^{\mu}f_{\lambda}^{\,\,\,\nu}-\partial_{\mu}f\partial^{\mu}f-\partial_{\mu}f_{\,\,\,\nu}^{\mu}\partial^{\lambda}f_{\lambda}^{\,\,\,\nu}+2\partial_{\mu}f\partial^{\lambda}f_{\lambda}^{\,\,\,\mu}\right].\label{eq:S_kin TEGR}
\end{equation}
We could now proceed as in the last section, i. e. Fourier transform
and then expand in the spin projection operators. However, in fact
we have already done this calculation: Comparing (\ref{eq:super metric flat})
with (\ref{eq:excitation form linear}), we see that $T_{\,\,\,\rho\sigma}^{\kappa}G_{\lambda\,\,\,\,\,\,\,\kappa}^{\,\,\,\mu\nu\,\,\,\rho\sigma}$
equals (\ref{eq:excitation form linear}) up to an overall factor.
This implies that (\ref{eq:S_kin TEGR}) produces the correct kinetic
term of Einstein gravity and hence also the correct propagator. This
was of course expected since we know that TEGR is perturbatively equivalent
to Einstein gravity. For the interaction terms the story will become
more non-trivial.

\subsection{Cubic vertex}

To calculate the cubic vertex , we need to expand both $G_{\lambda\,\,\,\,\,\,\,\kappa}^{\,\,\,\mu\nu\,\,\,\rho\sigma}$
and $e$ to first order in $f_{\,\,\,\mu}^{\lambda}.$ Let us start
with the super-metric:
\begin{align}
G_{\lambda\,\,\,\,\,\,\,\kappa}^{\,\,\,\mu\nu\,\,\,\rho\sigma}\mid_{f^{1}}= & -\eta_{\lambda\kappa}f^{\mu\rho}\eta^{\nu\sigma}-\eta_{\lambda\kappa}\eta^{\mu\rho}f^{\nu\sigma}-f_{\lambda}^{\,\,\,\sigma}\delta_{\kappa}^{\,\,\,\nu}\eta^{\mu\rho}-\delta_{\lambda}^{\,\,\,\sigma}f_{\kappa}^{\,\,\,\nu}\eta^{\mu\rho}\nonumber \\
 & -2\delta_{\lambda}^{\,\,\,\sigma}\delta_{\kappa}^{\,\,\,\nu}f^{\mu\rho}+2f_{\lambda}^{\,\,\,\nu}\delta_{\kappa}^{\,\,\,\sigma}\eta^{\mu\rho}+2\delta_{\lambda}^{\,\,\,\nu}f_{\kappa}^{\,\,\,\sigma}\eta^{\mu\rho}+4\delta_{\lambda}^{\,\,\,\nu}\delta_{\kappa}^{\,\,\,\sigma}f^{\mu\rho}.
\end{align}
Multiplying with $T_{\,\,\,\rho\sigma}^{\kappa}$ yields
\begin{align}
T_{\,\,\,\rho\sigma}^{\kappa}G_{\lambda\,\,\,\,\,\,\,\kappa}^{\,\,\,\mu\nu\,\,\,\rho\sigma}\mid_{f^{1}}= & -\partial_{\rho}f_{\lambda}^{\,\,\,\nu}f^{\mu\rho}+\partial^{\nu}f_{\kappa\rho}f^{\mu\rho}-\partial^{\mu}f_{\lambda\rho}f^{\nu\rho}+\partial_{\rho}f_{\lambda}^{\,\,\,\mu}f^{\nu\rho}-\partial^{\mu}f_{\rho}^{\,\,\,\nu}f_{\lambda}^{\,\,\,\rho}+\partial_{\rho}f^{\mu\nu}f_{\lambda}^{\,\,\,\rho}\nonumber \\
 & -\partial^{\mu}f_{\lambda}^{\,\,\,\kappa}f_{\kappa}^{\,\,\,\nu}+\partial_{\lambda}f^{\kappa\mu}f_{\kappa}^{\,\,\,\nu}-2\partial_{\rho}f_{\lambda}^{\,\,\,\nu}f^{\mu\rho}+2\partial_{\lambda}f_{\rho}^{\,\,\,\nu}f^{\mu\rho}+2\partial^{\mu}ff_{\lambda}^{\,\,\,\nu}-2\partial_{\kappa}f^{\kappa\mu}f_{\lambda}^{\,\,\,\nu}\nonumber \\
 & +2\partial^{\mu}f_{\,\,\,\rho}^{\kappa}f_{\kappa}^{\,\,\,\rho}\delta_{\lambda}^{\,\,\,\nu}-2\partial_{\rho}f^{\kappa\mu}f_{\kappa}^{\,\,\,\rho}\delta_{\lambda}^{\,\,\,\nu}+4\partial_{\rho}ff^{\mu\rho}\delta_{\lambda}^{\,\,\,\nu}-4\partial_{\kappa}f_{\,\,\,\rho}^{\kappa}f^{\mu\rho}\delta_{\lambda}^{\,\,\,\nu}.
\end{align}
We want to decompose this expression now as in (\ref{eq:c_decomposition}),
but first it needs to be antisymmetrized in $\mu$ and $\nu$. The
result is 
\begin{align}
T_{\,\,\,\rho\sigma}^{\kappa}G_{\lambda\,\,\,\,\,\,\,\kappa}^{\,\,\,\mu\nu\,\,\,\rho\sigma}\mid_{f^{1}}= & -2\left(\partial_{\rho}f_{\lambda}^{\,\,\,\nu}f^{\mu\rho}-\partial_{\rho}f_{\lambda}^{\,\,\,\mu}f^{\nu\rho}\right)+\frac{3}{2}\left(\partial^{\nu}f_{\lambda\rho}f^{\mu\rho}-\partial^{\mu}f_{\lambda\rho}f^{\nu\rho}\right)-\frac{1}{2}\left(\partial^{\mu}f_{\rho}^{\,\,\,\nu}f_{\lambda}^{\,\,\,\rho}-\partial^{\nu}f_{\rho}^{\,\,\,\mu}f_{\lambda}^{\,\,\,\rho}\right)\nonumber \\
 & -\frac{1}{2}\left(\partial_{\lambda}f^{\kappa\mu}f_{\kappa}^{\,\,\,\nu}-\partial_{\lambda}f^{\kappa\nu}f_{\kappa}^{\,\,\,\mu}\right)+\partial^{\mu}ff_{\lambda}^{\,\,\,\nu}-\partial^{\nu}ff_{\lambda}^{\,\,\,\mu}-\partial_{\kappa}f^{\kappa\mu}f_{\lambda}^{\,\,\,\nu}+\partial_{\kappa}f^{\kappa\nu}f_{\lambda}^{\,\,\,\mu}\nonumber \\
 & +\partial^{\mu}f_{\,\,\,\rho}^{\kappa}f_{\kappa}^{\,\,\,\rho}\delta_{\lambda}^{\,\,\,\nu}-\partial^{\nu}f_{\,\,\,\rho}^{\kappa}f_{\kappa}^{\,\,\,\rho}\delta_{\lambda}^{\,\,\,\mu}-\partial_{\rho}f^{\kappa\mu}f_{\kappa}^{\,\,\,\rho}\delta_{\lambda}^{\,\,\,\nu}+\partial_{\rho}f^{\kappa\nu}f_{\kappa}^{\,\,\,\rho}\delta_{\lambda}^{\,\,\,\mu}\nonumber \\
 & +2\left(\partial_{\rho}ff^{\mu\rho}\delta_{\lambda}^{\,\,\,\nu}-\partial_{\rho}ff^{\nu\rho}\delta_{\lambda}^{\,\,\,\mu}\right)-2\left(\partial_{\kappa}f_{\,\,\,\rho}^{\kappa}f^{\mu\rho}\delta_{\lambda}^{\,\,\,\nu}-\partial_{\kappa}f_{\,\,\,\rho}^{\kappa}f^{\nu\rho}\delta_{\lambda}^{\,\,\,\mu}\right).
\end{align}
Now we can read off the coefficients $\tilde{c}_{i}$ to be\footnote{We denote the coefficients stemming from the expansion of the super-metric
by $\tilde{c}_{i}$. } 
\begin{align*}
 & \tilde{c}_{1}=0,\,\,\,\,\,\,\,\,\,\,\,\,\tilde{c}_{2}=1,\,\,\,\,\,\,\,\,\,\,\,\,\tilde{c}_{3}=0,\,\,\,\,\,\,\,\,\,\,\,\,\tilde{c}_{4}=1,\,\,\,\,\,\,\,\,\,\,\,\,\tilde{c}_{5}=-\frac{1}{2},\,\,\,\,\,\,\,\,\,\,\,\,\tilde{c}_{6}=-\frac{3}{2},\,\,\,\,\,\,\,\,\,\,\,\,\,\tilde{c}_{7}=\frac{1}{2}\\
 & \tilde{c}_{8}=1,\,\,\,\,\,\,\,\,\,\,\,\,\tilde{c}_{9}=-2,\,\,\,\,\,\,\,\,\,\tilde{c}_{10}=0,\,\,\,\,\,\,\,\,\,\,\tilde{c}_{11}=2,\,\,\,\,\,\,\,\,\,\:\tilde{c}_{12}=-2,\,\,\,\,\,\,\,\,\,\,\,\,\tilde{c}_{13}=-1.
\end{align*}
The correlated coefficients $\tilde{d}_{i}$ for the full cubic vertex
are readily calculated using (\ref{eq:c-d relation}) and read 
\begin{align*}
 & \tilde{d}_{1}=0,\,\,\,\,\,\,\,\,\,\,\,\,\tilde{d}_{2}=0,\,\,\,\,\,\,\,\,\,\,\,\,\tilde{d}_{3}=0,\,\,\,\,\,\,\,\,\,\,\,\,\tilde{d}_{4}=0,\,\,\,\,\,\,\,\,\,\,\,\,\tilde{d}_{5}=0,\,\,\,\,\,\,\,\,\,\,\,\,\tilde{d}_{6}=-4,\,\,\,\,\,\,\,\,\,\,\,\,\,\tilde{d_{7}}=4,\,\,\,\,\,\,\,\,\,\,\,\,\,\tilde{d}_{8}=4\\
 & \tilde{d}_{9}=-4,\,\,\,\,\,\,\,\,\,\tilde{d}_{10}=-8,\,\,\,\,\,\,\,\,\,\tilde{d}_{11}=8,\,\,\,\,\,\,\,\,\,\,\tilde{d}_{12}=4,\,\,\,\,\,\,\,\,\,\:\tilde{d}_{13}=-4,\,\,\,\,\,\,\,\,\,\,\,\,\tilde{d}_{14}=4,\,\,\,\,\,\,\,\,\,\,\,\,\tilde{d}_{15}=-4,\,\,\,\,\,\,\,\,\tilde{d}_{16}=0.
\end{align*}

The contribution from the volume element is much simpler, we just
use 
\begin{equation}
e\mid_{f^{1}}=f,
\end{equation}
which has to be multiplied with the integrand in (\ref{eq:S_kin TEGR}).
The result can be expanded as in (\ref{eq:general 3-vertex}), yielding
the coefficients $\hat{d}_{i}:$\footnote{We denote the coefficients stemming from the expansion of the volume
element by $\hat{d}_{i}$.} 
\begin{equation}
\hat{d_{1}}=4,\,\,\,\,\,\,\,\,\,\,\,\,\hat{d}_{2}=-2,\,\,\,\,\,\,\,\,\,\,\,\,\,\hat{d}_{3}=-2,\,\,\,\,\,\,\,\,\,\,\,\,\,\hat{d}_{4}=2,\,\,\,\,\,\,\,\,\,\,\,\,\,\,\hat{d}_{5}=-2
\end{equation}
with all others vanishing. The full vertex is determined by the total
coefficients $d_{i}=\tilde{d}_{i}+\hat{d}_{i}$ which read 
\begin{align}
 & d_{1}=4,\,\,\,\,\,\,\,\,\,\,\,\,d_{2}=-2,\,\,\,\,\,\,\,\,\,\,\,\,d_{3}=-2,\,\,\,\,\,\,\,\,\,\,\,\,d_{4}=2,\,\,\,\,\,\,\,\,\,\,\,\,d_{5}=-2,\,\,\,\,\,\,\,\,\,\,\,\,d_{6}=-4,\,\,\,\,\,\,\,\,\,\,\,\,\,d_{7}=4,\,\,\,\,\,\,\,\,\,\,\,\,\,d_{8}=4\nonumber \\
 & d_{9}=-4,\,\,\,\,\,\,\,\,\,d_{10}=-8,\,\,\,\,\,\,\,\,\,d_{11}=8,\,\,\,\,\,\,\,\,\,\,d_{12}=4,\,\,\,\,\,\,\,\,\,\:d_{13}=-4,\,\,\,\,\,\,\,\,\,\,\,\,d_{14}=4,\,\,\,\,\,\,\,\,\,\,\,\,d_{15}=-4,\,\,\,\,\,\,\,\,d_{16}=0.\label{eq:TEGR d-values}
\end{align}
It is important to note this is not the standard three-graviton vertex.
Although TEGR is perturbatively equivalent to GR, it is formulated
in the vielbein formalism and uses a different graviton field. This
field redefinition will change the form of the vertices but leave
on-shell amplitudes invariant. 

\subsection{Quartic vertex}

To compute the full, gauge invariant 4-point amplitude, we also need
the contribution of the elementary quartic vertex. This means, we
need to expand the super-metric and the determinant $e$ to second
order in $f_{\,\,\,\mu}^{\lambda}.$ Let us start with the super-metric:
We use (\ref{eq:inverse vielbein}) as well as 
\begin{equation}
g^{\mu\nu}=\eta^{\mu\nu}-2f^{\mu\nu}+3f^{\mu\rho}f_{\rho}^{\,\,\,\nu}+\mathcal{O}\left(f^{3}\right)
\end{equation}
to get 
\begin{align}
G_{\alpha\,\,\,\,\,\,\,\beta}^{\,\,\,\mu\nu\,\,\,\rho\sigma}\mid_{f^{2}}= & 2\eta_{\alpha\beta}f^{\mu\rho}f^{\nu\sigma}+\frac{3}{2}\eta_{\alpha\beta}f^{\mu\lambda}f_{\lambda}^{\,\,\,\rho}\eta^{\nu\sigma}+\frac{3}{2}\eta_{\alpha\beta}\eta^{\mu\rho}f^{\nu\lambda}f_{\lambda}^{\,\,\,\sigma}\nonumber \\
 & +f_{\alpha}^{\,\,\,\sigma}f_{\beta}^{\,\,\,\nu}\eta^{\mu\rho}+2f_{\alpha}^{\,\,\,\sigma}\delta_{\beta}^{\,\,\,\nu}f^{\mu\rho}+2\delta_{\alpha}^{\,\,\,\sigma}f_{\beta}^{\,\,\,\nu}f^{\mu\rho}\nonumber \\
 & +f_{\alpha}^{\,\,\,\lambda}f_{\lambda}^{\,\,\,\sigma}\delta_{\beta}^{\,\,\,\nu}\eta^{\mu\rho}+\delta_{\alpha}^{\,\,\,\sigma}f_{\beta}^{\,\,\,\lambda}f_{\lambda}^{\,\,\,\nu}\eta^{\mu\rho}+3\delta_{\alpha}^{\,\,\,\sigma}\delta_{\beta}^{\,\,\,\nu}f^{\mu\lambda}f_{\lambda}^{\,\,\,\rho}\nonumber \\
 & -2f_{\alpha}^{\,\,\,\nu}f_{\beta}^{\,\,\,\sigma}\eta^{\mu\rho}-4f_{\alpha}^{\,\,\,\nu}\delta_{\beta}^{\,\,\,\sigma}f^{\mu\rho}-4\delta_{\alpha}^{\,\,\,\nu}f_{\beta}^{\,\,\,\sigma}f^{\mu\rho}\nonumber \\
 & -2f_{\alpha}^{\,\,\,\lambda}f_{\lambda}^{\,\,\,\nu}\delta_{\beta}^{\,\,\,\sigma}\eta^{\mu\rho}-2\delta_{\alpha}^{\,\,\,\nu}f_{\beta}^{\,\,\,\lambda}f_{\lambda}^{\,\,\,\sigma}\eta^{\mu\rho}-6\delta_{\alpha}^{\,\,\,\nu}\delta_{\beta}^{\,\,\,\sigma}f^{\mu\lambda}f_{\lambda}^{\,\,\,\rho}.
\end{align}
Multiplying with the two torsion tensors yields the already quite
lengthy expression
\begin{align}
 & T_{\,\,\,\mu\nu}^{\alpha}T_{\,\,\,\rho\sigma}^{\beta}G_{\alpha\,\,\,\,\,\,\,\beta}^{\,\,\,\mu\nu\,\,\,\rho\sigma}\mid_{f^{2}}=\nonumber \\
 & \,\,\,\,\,2\left(\partial_{\mu}f_{\,\,\,\nu}^{\alpha}-\partial_{\nu}f_{\,\,\,\mu}^{\alpha}\right)\left(\partial_{\rho}f_{\alpha\sigma}-\partial_{\sigma}f_{\alpha\rho}\right)f^{\mu\rho}f^{\nu\sigma}+\frac{3}{2}\left(\partial_{\mu}f_{\,\,\,}^{\alpha\sigma}-\partial^{\sigma}f_{\,\,\,\mu}^{\alpha}\right)\left(\partial_{\rho}f_{\alpha\sigma}-\partial_{\sigma}f_{\alpha\rho}\right)f^{\mu\lambda}f_{\lambda}^{\,\,\,\rho}\nonumber \\
 & +\frac{3}{2}\left(\partial_{\mu}f_{\,\,\,\nu}^{\alpha}-\partial_{\nu}f_{\,\,\,\mu}^{\alpha}\right)\left(\partial^{\mu}f_{\alpha\sigma}-\partial_{\sigma}f_{\alpha}^{\,\,\,\mu}\right)f^{\nu\lambda}f_{\lambda}^{\,\,\,\sigma}+\left(\partial_{\mu}f_{\,\,\,\nu}^{\alpha}-\partial_{\nu}f_{\,\,\,\mu}^{\alpha}\right)\left(\partial^{\mu}f_{\,\,\,\sigma}^{\beta}-\partial_{\sigma}f^{\beta\mu}\right)f_{\alpha}^{\,\,\,\sigma}f_{\beta}^{\,\,\,\nu}\nonumber \\
 & +2\left(\partial_{\mu}f_{\,\,\,\nu}^{\alpha}-\partial_{\nu}f_{\,\,\,\mu}^{\alpha}\right)\left(\partial_{\rho}f_{\,\,\,\sigma}^{\nu}-\partial_{\sigma}f_{\,\,\,\rho}^{\nu}\right)f_{\alpha}^{\,\,\,\sigma}f^{\mu\rho}+2\left(\partial_{\mu}f_{\,\,\,\nu}^{\alpha}-\partial_{\nu}f_{\,\,\,\mu}^{\alpha}\right)\left(\partial_{\rho}f_{\,\,\,\alpha}^{\beta}-\partial_{\alpha}f_{\,\,\,\rho}^{\beta}\right)f_{\beta}^{\,\,\,\nu}f^{\mu\rho}\nonumber \\
 & +\left(\partial_{\mu}f_{\,\,\,\nu}^{\alpha}-\partial_{\nu}f_{\,\,\,\mu}^{\alpha}\right)\left(\partial^{\mu}f_{\,\,\,\sigma}^{\nu}-\partial_{\sigma}f_{\,\,\,}^{\nu\mu}\right)f_{\alpha}^{\,\,\,\lambda}f_{\lambda}^{\,\,\,\sigma}+\left(\partial_{\mu}f_{\,\,\,\nu}^{\alpha}-\partial_{\nu}f_{\,\,\,\mu}^{\alpha}\right)\left(\partial^{\mu}f_{\,\,\,\alpha}^{\beta}-\partial_{\alpha}f_{\,\,\,}^{\beta\mu}\right)f_{\beta}^{\,\,\,\lambda}f_{\lambda}^{\,\,\,\nu}\nonumber \\
 & +3\left(\partial_{\mu}f_{\,\,\,\nu}^{\alpha}-\partial_{\nu}f_{\,\,\,\mu}^{\alpha}\right)\left(\partial_{\rho}f_{\,\,\,\alpha}^{\nu}-\partial_{\alpha}f_{\,\,\,\rho}^{\nu}\right)f^{\mu\lambda}f_{\lambda}^{\,\,\,\rho}-2\left(\partial_{\mu}f_{\,\,\,\nu}^{\alpha}-\partial_{\nu}f_{\,\,\,\mu}^{\alpha}\right)\left(\partial^{\mu}f_{\,\,\,\sigma}^{\beta}-\partial_{\sigma}f^{\beta\mu}\right)f_{\alpha}^{\,\,\,\nu}f_{\beta}^{\,\,\,\sigma}\nonumber \\
 & -4\left(\partial_{\mu}f_{\,\,\,\nu}^{\alpha}-\partial_{\nu}f_{\,\,\,\mu}^{\alpha}\right)\left(\partial_{\rho}f-\partial_{\sigma}f_{\,\,\,\rho}^{\sigma}\right)f_{\alpha}^{\,\,\,\nu}f^{\mu\rho}-4\left(\partial_{\mu}f-\partial_{\nu}f_{\,\,\,\mu}^{\nu}\right)\left(\partial_{\rho}f_{\,\,\,\sigma}^{\beta}-\partial_{\sigma}f_{\,\,\,\rho}^{\beta}\right)f_{\beta}^{\,\,\,\sigma}f^{\mu\rho}\nonumber \\
 & -2\left(\partial_{\mu}f_{\,\,\,\nu}^{\alpha}-\partial_{\nu}f_{\,\,\,\mu}^{\alpha}\right)\left(\partial^{\mu}f-\partial_{\sigma}f^{\sigma\mu}\right)f_{\alpha}^{\,\,\,\lambda}f_{\lambda}^{\,\,\,\nu}-2\left(\partial_{\mu}f-\partial_{\nu}f_{\,\,\,\mu}^{\nu}\right)\left(\partial^{\mu}f_{\,\,\,\sigma}^{\beta}-\partial_{\sigma}f^{\beta\mu}\right)f_{\beta}^{\,\,\,\lambda}f_{\lambda}^{\,\,\,\sigma}\nonumber \\
 & -6\left(\partial_{\mu}f-\partial_{\nu}f_{\,\,\,\mu}^{\nu}\right)\left(\partial_{\rho}f-\partial_{\sigma}f_{\,\,\,\rho}^{\sigma}\right)f^{\mu\lambda}f_{\lambda}^{\,\,\,\rho}.
\end{align}
This can be simplified though: If are only interested in the on-shell
4-point amplitude, we can omit all terms containing traces or divergences
of $f_{\mu\nu}$, since they will be killed by the on-shell conditions
anyway. This eliminates the last three lines and the rest can be collected
together to yield
\begin{align}
\tilde{V_{4}}^{on-shell}= & \,\,\,8\partial_{\mu}f_{\,\,\,\nu}^{\alpha}\partial_{\rho}f_{\alpha\sigma}f^{\mu\rho}f^{\nu\sigma}-7\partial_{\mu}f_{\,\,\,\nu}^{\alpha}\partial_{\sigma}f_{\alpha\rho}f^{\mu\rho}f^{\nu\sigma}+6\partial_{\mu}f^{\alpha\sigma}\partial_{\rho}f_{\alpha\sigma}f_{\,\,\,\lambda}^{\mu}f^{\lambda\rho}\nonumber \\
 & -12\partial_{\mu}f^{\alpha\sigma}\partial_{\sigma}f_{\alpha\rho}f_{\,\,\,\lambda}^{\mu}f^{\lambda\rho}+5\partial^{\sigma}f_{\,\,\,\mu}^{\alpha}\partial_{\sigma}f_{\alpha\rho}f_{\,\,\,\lambda}^{\mu}f^{\lambda\rho}\nonumber \\
 & +\partial_{\mu}f_{\,\,\,\nu}^{\alpha}\partial^{\mu}f_{\,\,\,\sigma}^{\beta}f_{\alpha}^{\,\,\,\sigma}f_{\beta}^{\,\,\,\nu}-2\partial_{\mu}f_{\,\,\,\nu}^{\alpha}\partial_{\sigma}f^{\beta\mu}f_{\alpha}^{\,\,\,\sigma}f_{\beta}^{\,\,\,\nu}+\partial_{\nu}f_{\,\,\,\mu}^{\alpha}\partial^{\mu}f_{\,\,\,\sigma}^{\nu}f_{\alpha\lambda}f^{\lambda\sigma}\nonumber \\
 & -2\partial_{\mu}f_{\,\,\,\nu}^{\alpha}\partial^{\mu}f_{\,\,\,\sigma}^{\beta}f_{\alpha}^{\,\,\,\nu}f_{\beta}^{\,\,\,\sigma}+4\partial_{\mu}f_{\,\,\,\nu}^{\alpha}\partial_{\sigma}f^{\beta\mu}f_{\alpha}^{\,\,\,\nu}f_{\beta}^{\,\,\,\sigma}-2\partial_{\mu}f_{\,\,\,\nu}^{\alpha}\partial_{\sigma}f_{\alpha\rho}f^{\mu\nu}f^{\rho\sigma}.
\end{align}

Now let us turn to the contribution from the volume element: In principle,
there should be terms where both the super-metric and the determinant
are expanded to first order. However, the first order expansion of
$e$ is just $f$, hence those terms will not contribute on-shell.
The second order term is just given by 
\begin{equation}
e\mid_{f^{2}}=\frac{1}{2}\left(f^{2}-f_{\mu\nu}f^{\mu\nu}\right),
\end{equation}
where we can immediately omit the $f^{2}$ and multiply with (\ref{eq:S_kin TEGR})
to obtain 
\begin{align*}
\hat{V}_{4}^{on-shell}=- & \partial_{\mu}f_{\,\,\,\nu}^{\lambda}\partial^{\mu}f_{\lambda}^{\,\,\,\nu}f_{\rho\sigma}f^{\rho\sigma}+\partial_{\nu}f_{\,\,\,\mu}^{\lambda}\partial^{\mu}f_{\lambda}^{\,\,\,\nu}f_{\rho\sigma}f^{\rho\sigma}
\end{align*}
 after omitting all terms that vanish on-shell. The total on-shell
four-vertex is just the sum 
\begin{equation}
V_{4}^{on-shell}=\tilde{V}_{4}^{on-shell}+\hat{V}_{4}^{on-shell},
\end{equation}
its coefficients in terms of (\ref{eq:most gen 4vert}) are 
\begin{align}
 & l_{1}=-1,\,\,\,\,\,\,\,\,\,l_{2}=5,\,\,\,\,\,\,\,\,\,l_{3}=-2,\,\,\,\,\,\,\,\,\,\,l_{4}=1,\,\,\,\,\,\,\,\,\,\,\,l_{5}=1,\,\,\,\,\,\,\,\,\,\,l_{6}=1,\,\,\,\,\,\,\,\,\,\,\,l_{7}=-12,\nonumber \\
 & l_{8}=4,\,\,\,\,\,\,\,\,\,\,\,\,\,l_{9}=-2,\,\,\,\,\,\,\,\,\,\,\,l_{10}=6,\,\,\,\,\,\,\,\,\,\,\,\,l_{11}=-2,\,\,\,\,\,\,\,\,\,\,\,\,l_{12}=8,\,\,\,\,\,\,\,\,\,\,\,l_{13}=-7.
\end{align}
This concludes the section on perturbative teleparallel gravity.

\section{Calculation of on-shell amplitudes}

We are now in the position to calculate tree-level on-shell amplitudes
using our bottom-up constructed double-copy Lagrangian and match them
with the known double-copy results. This will allow us to further
constrain the coefficients within the vertices. The results we get
can then be compared with what we know about teleparallel gravity.
The first step is to fix the relative normalizations of the Lagrangians:
We have seen in (\ref{eq:prop}) that our double-copy action with
the choice $\alpha=2$ is not canonically normalized, e. g. the on-shell
residue of the propagator is not one. While this choice is convenient
for the comparison with TEGR, we must keep track of the factor $\alpha+2=4$
in the following. Note that the teleparallel Lagrangian we used to
derive the propagator and the vertices, also differs from the conventional,
canonically normalized Einstein-Hilbert Lagrangian by a factor of
four:
\begin{equation}
\mathcal{L}_{TEGR}=e\,T_{\,\,\,\mu\nu}^{\lambda}T_{\,\,\,\rho\sigma}^{\kappa}G_{\lambda\,\,\,\,\,\,\,\kappa}^{\,\,\,\mu\nu\,\,\,\rho\sigma}=2\sqrt{-g}\stackrel{\circ}{R},\,\,\,\,\,\,\,\,\,\,\,\,\,\,\,\,,\mathcal{L}_{GR}=\frac{1}{2}\sqrt{-g}\stackrel{\circ}{R}.
\end{equation}
It is therefore convenient to define our double-copy prescription
with an additional factor of four for all amplitudes:
\begin{equation}
i\mathcal{M}^{DC}\stackrel{!}{=}4i\mathcal{A}^{gluon}\mid_{c_{i}\rightarrow n_{i}}
\end{equation}

\subsection{3-point amplitude}

We start with the on-shell 3-point amplitude. For real momenta, it
is generically zero for gluons as well as for gravitons. However,
by analytically continuing to complex momenta, we get non-vanishing
results that can be compared meaningfully. The 3-point gluon amplitude
only consists of the elementary 3-vertex contracted with three on-shell
states and therefore takes the simple form 
\begin{align}
i\mathcal{A}_{3}^{gluon} & =gf^{abc}\left(\left(p_{1}-p_{2}\right)_{\rho}\eta_{\mu\nu}+\left(p_{2}-p_{3}\right)_{\mu}\eta_{\nu\rho}+\left(p_{3}-p_{1}\right)_{\nu}\eta_{\mu\rho}\right)\epsilon_{1a}^{\mu}\epsilon_{2b}^{\nu}\epsilon_{3c}^{\rho}\nonumber \\
 & =gf^{abc}\Big(\left(\left(p_{1}\cdot\epsilon_{3c}\right)-\left(p_{2}\cdot\epsilon_{3c}\right)\right)\left(\epsilon_{1a}\cdot\epsilon_{2b}\right)+\left(\left(p_{2}\cdot\epsilon_{1a}\right)-\left(p_{3}\cdot\epsilon_{1a}\right)\right)\left(\epsilon_{2b}\cdot\epsilon_{3c}\right)\nonumber \\
 & \,\,\,\,\,\,\,\,\,\,\,\,\,\,\,\,\,\,\,\,\,\,\,\,\,\left(\left(p_{3}\cdot\epsilon_{2b}\right)-\left(p_{1}\cdot\epsilon_{2b}\right)\right)\left(\epsilon_{1a}\cdot\epsilon_{3c}\right)\Big).
\end{align}
There is no denominator and the relevant numerator $n$ that appears
in the double-copy relation can be identified simply by stripping
off the colour factor and reads 
\begin{align*}
n & =\Big(\left(\left(p_{1}\cdot\epsilon_{3}\right)-\left(p_{2}\cdot\epsilon_{3}\right)\right)\left(\epsilon_{1}\cdot\epsilon_{2}\right)+\left(\left(p_{2}\cdot\epsilon_{1}\right)-\left(p_{3}\cdot\epsilon_{1}\right)\right)\left(\epsilon_{2}\cdot\epsilon_{3}\right)+\left(\left(p_{3}\cdot\epsilon_{2}\right)-\left(p_{1}\cdot\epsilon_{2}\right)\right)\left(\epsilon_{1}\cdot\epsilon_{3}\right)\Big).
\end{align*}
The original relation, including the coupling $\kappa=2$ in the conventions
of \cite{Bern2024} reads 
\[
i\mathcal{M}_{3}=\frac{i}{2}\mathcal{A}_{3}^{gluon},
\]
hence we expect the double-copy amplitude to fulfill
\begin{equation}
i\mathcal{M}_{3}^{DC}=-2n^{2}.\label{eq:3-point DC rel}
\end{equation}
Let us compute $i\mathcal{M}_{3}^{DC}$: As explained in section \ref{subsec:Cubic-term},
only the terms corresponding to $d_{9}$, $d_{11},$$d_{15}$ and
$d_{16}$ in (\ref{eq:general 3-vertex}) contribute with the constraint
\begin{equation}
d_{9}+d_{11}+d_{15}+d_{16}=0,
\end{equation}
hence 
\begin{equation}
V_{3}^{on-shell}=d_{9}f_{\mu\nu}\partial^{\mu}f^{\rho\sigma}\partial^{\nu}f_{\rho\sigma}+d_{11}f_{\mu\nu}\partial^{\mu}f_{\rho\sigma}\partial^{\rho}f^{\nu\sigma}-\left(d_{9}+d_{11}+d_{16}\right)f_{\mu\nu}\partial_{\rho}f^{\mu\sigma}\partial^{\rho}f_{\sigma}^{\nu}+d_{16}f_{\mu\nu}\partial_{\rho}f^{\nu\sigma}\partial_{\sigma}f^{\mu\rho}.
\end{equation}
Fourier transforming and contracting with the external states yields
\begin{align}
-i\mathcal{M}_{3}^{DC}= & d_{9}\left(\epsilon_{1}\cdot p_{2}\right)\left(\epsilon_{1}\cdot p_{3}\right)\left(\epsilon_{2}\cdot\epsilon_{3}\right)^{2}+d_{11}\left(\epsilon_{1}\cdot p_{2}\right)\left(\epsilon_{1}\cdot\epsilon_{3}\right)\left(\epsilon_{2}\cdot p_{3}\right)\left(\epsilon_{2}\cdot\epsilon_{3}\right)-\nonumber \\
 & \left(d_{9}+d_{11}+d_{16}\right)\left(\epsilon_{1}\cdot p_{2}\right)\left(\epsilon_{1}\cdot\epsilon_{3}\right)\left(p_{2}\cdot p_{3}\right)\left(\epsilon_{2}\cdot\epsilon_{3}\right)+d_{16}\left(\epsilon_{1}\cdot\epsilon_{2}\right)\left(\epsilon_{1}\cdot\epsilon_{3}\right)\left(\epsilon_{3}\cdot p_{2}\right)\left(\epsilon_{2}\cdot p_{3}\right)+perm
\end{align}
where ``perm'' stands for the in total six permutations of the external
states. For generic states, the non-vanishing products $\epsilon_{i}\cdot\epsilon_{j}$
and $\epsilon_{i}\cdot p_{j}$ will be taken to be independent and
only subject to momentum conservation. This allows us to solve equation
(\ref{eq:3-point DC rel}) for the coefficients $d_{9}$, $d_{11}$
and $d_{16}$. The calculation was done using Mathematica Xact and
the result is 
\begin{equation}
d_{9}=-4,\,\,\,\,\,\,\,\,\,\,\,\,\,\,\,\,d_{16}=d_{11}-8.\label{eq:3-point results}
\end{equation}
One should notice that the $d_{15}$-term actually drops out of the
calculation completely. This is because the product $\left(p_{2}\cdot p_{3}\right)$,
after summing over all permutations, is just proportional to 
\begin{equation}
s+t+u=0.
\end{equation}
The same phenomenon will also occur in the quartic amplitude.

\subsection{4-point amplitude}

Let us turn now to the 4-point amplitude, which has a far less trivial
structure. First of all, we should observe that we cannot compare
the individual terms in (\ref{eq:4-point grav amp}): Different Lagrangians
of the same theory will give in general different expressions for
the individual diagrams, only the total amplitude is invariant. According
to section two the expected amplitude is given by 
\begin{equation}
i\mathcal{M}_{4}=4\left(\frac{n_{s}^{2}}{s}+\frac{n_{t}^{2}}{t}+\frac{n_{u}^{2}}{u}\right)\label{eq:DC amplitude}
\end{equation}
with
\begin{align}
n_{s}=- & \frac{1}{2}\Big(\left[\left(\epsilon_{1}\cdot\epsilon_{2}\right)p_{1}^{\mu}+2\left(\epsilon_{1}\cdot p_{2}\right)\epsilon_{2}^{\mu}-\left(\epsilon_{2}\cdot\epsilon_{1}\right)p_{2}^{\mu}-2\left(\epsilon_{2}\cdot p_{1}\right)\epsilon_{1}^{\mu}\right]\nonumber \\
 & \left[\left(\epsilon_{3}\cdot\epsilon_{4}\right)p_{3\mu}+2\left(\epsilon_{3}\cdot p_{4}\right)\epsilon_{4\mu}-\left(\epsilon_{4}\cdot\epsilon_{3}\right)p_{4\mu}-2\left(\epsilon_{4}\cdot p_{3}\right)\epsilon_{3\mu}\right]\nonumber \\
 & +s\left(\left(\epsilon_{1}\cdot\epsilon_{3}\right)\left(\epsilon_{2}\cdot\epsilon_{4}\right)-\left(\epsilon_{1}\cdot\epsilon_{4}\right)\left(\epsilon_{2}\cdot\epsilon_{3}\right)\right)\Big)
\end{align}
and likewise for $n_{t}$ and $n_{u}.$ 

On the other hand, to calculate $\mathcal{M}_{4}$ from the double-copy
Lagrangian, we first note that the $s$-channel diagram has the form
\begin{equation}
V_{3}\left(p_{1},\epsilon_{1};p_{2},\epsilon_{2}\right)^{\mu\nu}\Pi_{\mu\nu\rho\sigma}V_{3}\left(p_{3},\epsilon_{3};p_{4},\epsilon_{4}\right)^{\rho\sigma}.
\end{equation}
As we have argued (\ref{eq:prop}), the propagator we will use is
just the GR-propagator in Feynman gauge up to a normalization constant:
\begin{equation}
\Pi_{\mu\nu\rho\sigma}=\frac{1}{2\left(\alpha+2\right)p^{2}}\left(\eta_{\mu\rho}\eta_{\nu\sigma}+\eta_{\mu\sigma}\eta_{\nu\rho}-\eta_{\mu\nu}\eta_{\rho\sigma}\right)
\end{equation}
with $p=p_{1}+p_{2}$. The vertices in the form $V_{3}\left(p_{1},\epsilon_{1};p_{2},\epsilon_{2}\right)^{\mu\nu}$
can be obtained by Fourier transforming (\ref{eq:general 3-vertex})
and inserting two external states in all different combinations. In
general, all terms associated to $\mathscr{V}_{rel4}=\left\{ d_{4},d_{5},d_{6},d_{7},d_{9},d_{11},d_{12},d_{13},d_{15},d_{16}\right\} $
contribute to the $s$-channel diagram. We can therefore write, already
using (\ref{eq:3-point results}) and (\ref{eq:d_relations}), 
\begin{align}
V_{3}^ {}\left(p_{1},\epsilon_{1};p_{2},\epsilon_{2}\right)^{\mu\nu}=\, & \left(2d_{4}-2d_{12}\right)\left(\epsilon_{1}\cdot\epsilon_{2}\right)^{2}\frac{s}{2}\eta^{\mu\nu}-\left(2d_{4}+2d_{6}\right)\left(\epsilon_{1}\cdot\epsilon_{2}\right)\left(p_{1}\cdot\epsilon_{2}\right)\left(p_{2}\cdot\epsilon_{1}\right)\eta^{\mu\nu}\nonumber \\
 & -\left(8+2d_{13}\right)\left(\epsilon_{1}\cdot\epsilon_{2}\right)^{2}p_{1}^{\mu}p_{2}^{\nu}+8\left(p_{2}\cdot\epsilon_{1}\right)^{2}\epsilon_{2}^{\mu}\epsilon_{2}^{\nu}+8\left(p_{1}\cdot\epsilon_{2}\right)^{2}\epsilon_{1}^{\mu}\epsilon_{1}^{\nu}\nonumber \\
 & +\left(16-d_{11}+d_{6}+d_{12}+d_{13}\right)\left(\epsilon_{1}\cdot\epsilon_{2}\right)\left(p_{2}\cdot\epsilon_{1}\right)p_{1}^{\mu}\epsilon_{2}^{\nu}\nonumber \\
 & +\left(16-d_{11}+d_{6}+d_{12}+d_{13}\right)\left(\epsilon_{1}\cdot\epsilon_{2}\right)\left(p_{1}\cdot\epsilon_{2}\right)p_{2}^{\mu}\epsilon_{1}^{\nu}\nonumber \\
 & +\left(-d_{11}+d_{6}+d_{12}+d_{13}\right)\left(\epsilon_{1}\cdot\epsilon_{2}\right)\left(p_{2}\cdot\epsilon_{1}\right)p_{2}^{\mu}\epsilon_{2}^{\nu}\nonumber \\
 & +\left(-d_{11}+d_{6}+d_{12}+d_{13}\right)\left(\epsilon_{1}\cdot\epsilon_{2}\right)\left(p_{1}\cdot\epsilon_{2}\right)p_{1}^{\mu}\epsilon_{1}^{\nu}-16\left(p_{1}\cdot\epsilon_{2}\right)\left(p_{2}\cdot\epsilon_{1}\right)\epsilon_{1}^{\mu}\epsilon_{2}^{\nu}\nonumber \\
 & +\left(24-4d_{11}\right)\left(\epsilon_{1}\cdot\epsilon_{2}\right)\frac{s}{2}\epsilon_{1}^{\mu}\epsilon_{2}^{\nu}-d_{13}\left(\epsilon_{1}\cdot\epsilon_{2}\right)^{2}p_{1}^{\mu}p_{1}^{\nu}-d_{13}\left(\epsilon_{1}\cdot\epsilon_{2}\right)^{2}p_{2}^{\mu}p_{2}^{\nu}\label{eq:LVert}
\end{align}
and likewise for $V_{3}\left(p_{3},\epsilon_{3};p_{4},\epsilon_{4}\right)^{\rho\sigma}$.
Now, multiplying everything out yields a huge number of terms that
be handled nicely using Mathematica Xact. The $t$- and $u$-channel
diagram can be obtained by simple permutations of the external labels.

For the quartic interaction diagram we need to Fourier transform (\ref{eq:most gen 4vert})
and insert the external states in all 24 different combinations. The
result is 
\begin{align}
 & V_{4}\left(p_{1},\epsilon_{1};p_{2},\epsilon_{2};p_{3},\epsilon_{3};p_{4},\epsilon_{4}\right)\nonumber \\
=\, & l_{12}\left(p_{1}\cdot\epsilon_{3}\right)\left(\epsilon_{1}\cdot\epsilon_{2}\right)\left(\epsilon_{1}\cdot\epsilon_{4}\right)\left(p_{2}\cdot\epsilon_{3}\right)\left(\epsilon_{2}\cdot\epsilon_{4}\right)+l_{13}\left(p_{1}\cdot\epsilon_{3}\right)\left(\epsilon_{1}\cdot\epsilon_{2}\right)\left(\epsilon_{1}\cdot\epsilon_{4}\right)\left(p_{2}\cdot\epsilon_{4}\right)\left(\epsilon_{2}\cdot\epsilon_{3}\right)\nonumber \\
 & +l_{10}\left(p_{1}\cdot\epsilon_{3}\right)\left(\epsilon_{1}\cdot\epsilon_{2}\right)^{2}\left(p_{2}\cdot\epsilon_{4}\right)\left(\epsilon_{3}\cdot\epsilon_{4}\right)+l_{7}\left(p_{1}\cdot\epsilon_{3}\right)\left(\epsilon_{1}\cdot\epsilon_{2}\right)\left(\epsilon_{2}\cdot\epsilon_{4}\right)\left(p_{2}\cdot\epsilon_{1}\right)\left(\epsilon_{3}\cdot\epsilon_{4}\right)\nonumber \\
 & +\left(-l_{6}-l_{7}-l_{10}\right)\frac{s}{2}\left(\epsilon_{1}\cdot\epsilon_{2}\right)\left(\epsilon_{1}\cdot\epsilon_{3}\right)\left(\epsilon_{2}\cdot\epsilon_{4}\right)\left(\epsilon_{3}\cdot\epsilon_{4}\right)+\left(-l_{9}-l_{12}-l_{13}\right)\frac{s}{2}\left(\epsilon_{1}\cdot\epsilon_{3}\right)\left(\epsilon_{1}\cdot\epsilon_{4}\right)\left(\epsilon_{2}\cdot\epsilon_{4}\right)\left(\epsilon_{2}\cdot\epsilon_{3}\right)\nonumber \\
 & +l_{9}\left(p_{1}\cdot\epsilon_{2}\right)\left(\epsilon_{1}\cdot\epsilon_{3}\right)\left(\epsilon_{1}\cdot\epsilon_{4}\right)\left(p_{2}\cdot\epsilon_{3}\right)\left(\epsilon_{2}\cdot\epsilon_{4}\right)+l_{6}\left(p_{1}\cdot\epsilon_{2}\right)\left(\epsilon_{1}\cdot\epsilon_{3}\right)\left(\epsilon_{2}\cdot\epsilon_{4}\right)\left(p_{2}\cdot\epsilon_{1}\right)\left(\epsilon_{3}\cdot\epsilon_{4}\right)\nonumber \\
 & +\left(-l_{8}-l_{11}\right)\frac{s}{2}\left(\epsilon_{1}\cdot\epsilon_{3}\right)^{2}\left(\epsilon_{2}\cdot\epsilon_{4}\right)^{2}+l_{8}\left(p_{1}\cdot\epsilon_{2}\right)\left(\epsilon_{1}\cdot\epsilon_{3}\right)^{2}\left(p_{2}\cdot\epsilon_{4}\right)\left(\epsilon_{2}\cdot\epsilon_{4}\right)\nonumber \\
 & +l_{11}\left(p_{1}\cdot\epsilon_{3}\right)\left(\epsilon_{1}\cdot\epsilon_{2}\right)\left(\epsilon_{1}\cdot\epsilon_{3}\right)\left(p_{2}\cdot\epsilon_{4}\right)\left(\epsilon_{2}\cdot\epsilon_{4}\right)-l_{5}\frac{s}{2}\left(\epsilon_{1}\cdot\epsilon_{2}\right)^{2}\left(\epsilon_{3}\cdot\epsilon_{4}\right)^{2}\nonumber \\
 & +l_{5}\left(p_{1}\cdot\epsilon_{2}\right)\left(\epsilon_{1}\cdot\epsilon_{2}\right)^ {}\left(p_{2}\cdot\epsilon_{1}\right)\left(\epsilon_{3}\cdot\epsilon_{4}\right)^{2}+23\,perm,
\end{align}
already using (\ref{eq:l-relations}). The total amplitude is then
given by the sum of the three exchange diagrams and the contact diagram.
It is a lengthy expression containing polynomials of order five in
the variables $s$, $t$, $u,$ $p_{i}\cdot\epsilon_{j},$ $\epsilon_{i}\cdot\epsilon_{j}$.
Since the external particles are generic, we will assume those polynomials
are independent apart from momentum conservation. The coefficients
are quadratic in the $d_{i}$ and linear in the $l_{i}$. By direct
comparison with (\ref{eq:DC amplitude}) we can therefore derive the
following relations:\begin{equation}\label{eq:amp-red-1}
4d_{11}-l_{12}-24=0.
\end{equation}

\begin{equation}\label{eq:amp-red-2}
(d_4+d_6)(d_{11}-d_{12}-d_{13}+2d_4+d_6-8)=0.
\end{equation}

\begin{equation}\label{eq:amp-red-3}
\begin{aligned}
0={}&2d_{11}^2-4d_{11}d_{12}-3d_{11}d_{13}+6d_{11}d_4+2d_{11}d_6-28d_{11}\\
&+2d_{12}^2+d_{12}d_{13}-10d_{12}d_4-6d_{12}d_6+28d_{12}\\
&-d_{13}^2-4d_{13}d_4-3d_{13}d_6+16d_{13}+8d_4^2+6d_4d_6\\
&-48d_4-20d_6+16l_5-8l_8+96.
\end{aligned}
\end{equation}

\begin{equation}\label{eq:amp-red-4}
8d_{11}-2l_6+l_7+l_9-48=0.
\end{equation}

\begin{equation}\label{eq:amp-red-5}
\begin{aligned}
0={}&3d_{11}^2-2d_{11}d_{12}-2d_{11}d_{13}-2d_{11}d_6-8d_{11}\\
&-d_{12}^2-2d_{12}d_{13}-2d_{12}d_6+8d_{12}-d_{13}^2\\
&-2d_{13}d_6+8d_{13}-d_6^2+8d_6+16l_{11}+16l_{13}.
\end{aligned}
\end{equation}

\begin{equation}\label{eq:amp-red-6}
d_{11}-l_{10}-2=0.
\end{equation}

\begin{equation}\label{eq:amp-red-7}
\begin{aligned}
0={}&3d_{11}^2-2d_{11}d_{12}-2d_{11}d_{13}-2d_{11}d_6-40d_{11}\\
&-d_{12}^2-2d_{12}d_{13}-2d_{12}d_6+8d_{12}-d_{13}^2\\
&-2d_{13}d_6+8d_{13}-d_6^2+8d_6+16l_{11}-8l_7+8l_9+64.
\end{aligned}
\end{equation}

\begin{equation}\label{eq:amp-red-8}
4d_{11}+2l_{13}+l_7-l_9-8=0.
\end{equation}

\begin{equation}\label{eq:amp-red-9}
\begin{aligned}
0={}&3d_{11}^2-2d_{11}d_{12}-2d_{11}d_{13}-2d_{11}d_6-104d_{11}\\
&-d_{12}^2-2d_{12}d_{13}-2d_{12}d_6+8d_{12}-d_{13}^2\\
&-2d_{13}d_6+8d_{13}-d_6^2+8d_6+16l_{11}+16l_6\\
&-16l_7+448.
\end{aligned}
\end{equation}

\begin{equation}\label{eq:amp-red-10}
\begin{aligned}
0={}&2d_{11}^2-4d_{11}d_{12}-3d_{11}d_{13}+6d_{11}d_4+2d_{11}d_6-36d_{11}\\
&+2d_{12}^2+d_{12}d_{13}-10d_{12}d_4-6d_{12}d_6+28d_{12}\\
&-d_{13}^2-4d_{13}d_4-3d_{13}d_6+16d_{13}+8d_4^2+6d_4d_6\\
&-48d_4-20d_6+8l_{10}+16l_5-8l_8+112.
\end{aligned}
\end{equation}

\begin{equation}\label{eq:amp-red-11}
2d_{11}-l_{13}-l_6+l_9-20=0.
\end{equation}

\begin{equation}\label{eq:amp-red-12}
\begin{aligned}
0={}&4d_{11}^2-8d_{11}d_{12}-4d_{11}d_{13}+8d_{11}d_4-64d_{11}\\
&+8d_{12}^2+4d_{12}d_{13}-16d_{12}d_4+64d_{12}-d_{13}^2\\
&-4d_{13}d_4+24d_{13}+8d_4^2-64d_4-16l_{11}+32l_5-16l_8+240.
\end{aligned}
\end{equation}

\begin{equation}\label{eq:amp-red-13}
d_{11}^2-12d_{11}-4l_{10}+8l_{12}+8l_{13}-4l_6-4l_7+8l_9+20=0.
\end{equation}

The second equation shows that we can distinguish two branches:
\begin{equation}
d_{6}=-2d_{4}-d_{11}+d_{12}+d_{13}+8,\,\,\,\,\,\,\,\,\,\,\,\,\,\,\,\,\,d'_{6}=-d'_{4}
\end{equation}
Within each branch, we can eliminate the $l_{i}$ to derive a quadratic
equation for $d_{4}$ :
\begin{equation}
0=60d_{4}^{2}+\left(-120d_{12}-60d_{13}+80d_{16}\right)d_{4}+60d_{12}^{2}+60d_{12}d_{13}-80d_{12}d_{16}+15d_{13}^{2}-40d_{13}d_{16}+36d_{16}^{2},
\end{equation}
\begin{equation}
0=15d'{}_{4}^{2}+\left(10d'_{16}-30d'_{12}\right)d'_{4}+15d'{}_{12}^{2}-10d'_{12}d'_{16}+11d'{}_{16}^{2}.
\end{equation}
The discriminants are given by $-2240d_{16}^{2}$ and $-560d'{}_{16}^{2}$,
respectively, hence if we focus on real solutions only, in both cases
we have to demand 
\begin{equation}
d_{16}=0.
\end{equation}
Together with the previously derived relations we can write the full
list $\mathscr{V}_{rel4}$ all in terms of two free variables $d_{4}$
and $d_{12}$ or $d'_{4}$ and $d'_{13}$ as 
\begin{align}
 & d_{5}=-d_{4},\,\,\,\,\,\,\,\,\,\,\,\,\,\,\,d_{6}=-d_{12},\,\,\,\,\,\,\,\,\,\,\,\,\,\,\,d_{7}=2d_{12}-2d_{4},\,\,\,\,\,\,\,\,\,\,\,\,\,\,\,\,\,d_{9}=-4,\nonumber \\
 & d_{11}=8,\,\,\,\,\,\,\,\,\,\,\,\,\,\,\,\,d_{13}=2d_{4}-2d_{12},\,\,\,\,\,\,\,\,\,\,\,\,\,\,\,\,d_{15}=-4,\,\,\,\,\,\,\,\,\,\,\,\,\,\,\,\,\,\,d_{16}=0,
\end{align}

\begin{align}
 & d'_{5}=-d'_{4},\,\,\,\,\,\,\,\,\,\,\,\,\,\,\,d'_{6}=-d'_{4},\,\,\,\,\,\,\,\,\,\,\,\,\,\,\,d'_{7}=-d'_{13},\,\,\,\,\,\,\,\,\,\,\,\,\,\,\,\,\,d'_{9}=-4,\nonumber \\
 & d'_{11}=8,\,\,\,\,\,\,\,\,\,\,\,\,\,\,\,\,\,d'_{12}=d'_{4},\,\,\,\,\,\,\,\,\,\,\,\,\,\,\,\,d'_{15}=-4,\,\,\,\,\,\,\,\,\,\,\,\,\,\,\,\,\,\,d'_{16}=0.
\end{align}
A quick comparison with (\ref{eq:TEGR d-values}) shows that the first
branch is compatible with perturbative teleparallel gravity for $d_{4}=2$
and $d_{12}=4$, as we intended to show. The solutions for the $l_{i}$
are also dependent on two free parameters and can be expressed in
terms of $l_{8}$ and $l_{9}$:
\begin{align}
 & l_{1}=-\frac{\left(d_{12}^{2}-2d_{12}d_{4}-4d_{12}+d_{4}^{2}+4d_{4}+2l_{8}\right)}{4},\,\,\,\,\,\,\,\,\,\,\,\,\,\,\,\,\,l_{2}=-2l_{9}+1,\nonumber \\
 & l_{3}=-l_{8}-\frac{\left(d_{12}^{2}-2d_{12}d_{4}-4d_{12}+d_{4}^{2}+4d_{4}-4\right)}{4},\,\,\,\,\,\,\,\,\,\,\,\,\,\,\,\,\,l_{4}=-l_{9}-1\nonumber \\
 & l_{5}=\frac{\left(d_{12}^{2}-2d_{12}d_{4}-4d_{12}+d_{4}^{2}+4d_{4}+2l_{8}\right)}{4},\,\,\,\,\,\,\,\,\,\,\,\,\,\,\,\,\,\,l_{6}=l_{9}+3,\,\,\,\,\,\,\,\,\,\,\,\,\,\,\,\,l_{7}=l_{9}-10\nonumber \\
 & l_{10}=6,\,\,\,\,\,\,\,\,\,\,\,\,\,\,l_{11}=\frac{\left(d_{12}^{2}-2d_{12}d_{4}-4d_{12}+d_{4}^{2}+4d_{4}-4\right)}{4},\,\,\,\,\,\,\,\,\,\,\,\,\,\,\,\,\,\,l_{12}=8,\,\,\,\,\,\,\,\,\,\,\,\,\,\,\,\,\,\,\,l_{13}=-7,
\end{align}
\begin{align}
 & l'_{1}=-\frac{\left(d'{}_{13}^{2}+8d'_{13}+8l'_{8}\right)}{16},\,\,\,\,\,\,\,\,\,\,\,\,\,\,\,\,\,l'_{2}=-2l'_{9}+1,\nonumber \\
 & l'_{3}=-l'_{8}-\frac{\left(d'{}_{13}^{2}+8d'_{13}-16\right)}{16},\,\,\,\,\,\,\,\,\,\,\,\,\,\,\,\,\,l'_{4}=-l'_{9}-1\nonumber \\
 & l'_{5}=\frac{\left(d'{}_{13}^{2}+8d'_{13}+8l'_{8}\right)}{16}\,\,\,\,\,\,\,\,\,\,\,\,\,\,\,\,\,\,l'_{6}=l'_{9}+3,\,\,\,\,\,\,\,\,\,\,\,\,\,\,\,\,l'_{7}=l'_{9}-10\nonumber \\
 & l'_{10}=6,\,\,\,\,\,\,\,\,\,\,\,\,\,\,l'_{11}=\frac{\left(d'{}_{13}^{2}+8d'_{13}-16\right)}{16},\,\,\,\,\,\,\,\,\,\,\,\,\,\,\,\,\,\,l'_{12}=8,\,\,\,\,\,\,\,\,\,\,\,\,\,\,\,\,\,\,\,l'_{13}=-7,
\end{align}
Again we can see immediately that for the first branch, the choice
$l_{8}=4$, $l_{9}=-2$ exactly reproduces the values derived from
teleparallel gravity.

\subsection{Higher amplitudes}

By computing the on-shell three- and four-point amplitude and matching
it with the double copy results, we have constrained the coefficients
of the vertices to a high extent: For the cubic vertex, the teleparallel
values were derived up to two real parameters and one discrete choice,
whileas for the on-shell quartic vertex another two free parameters
remained. It is a natural question to ask whether the vertices can
be fully determined by computing higher amplitudes. Although a complete
calculation of the five-point amplitude for instance goes beyond the
scope of this paper, we will give some qualitative arguments. 

In principle, with the propagator given in its current form, it might
be possible to determine all 16 $d_{i}$. At higher order they will
contribute: For instance the $d_{3}$-term contains three traces,
so it can only be non-vanishing if all three legs go into propagators.
In the six- and higher point amplitudes, there are diagrams where
this happens, so those coefficients will certainly appear in the calculation.
This does not imply that they all can be determined separately though,
as explained below. 

The second question is about the discrete branch choice and the two
free parameters: The branch split arises from a quadratic equation
involving $d$-coefficients only. At higher orders, we will get higher
polynomial conditions in the $d_{i}$, for instance from diagrams
with cubic vertices only. One can expect them to be strong enough
to select the correct branch. The question about the two free parameters
is a bit more subtle: Look for instance at the $d_{4}$- and the $d_{12}$-term:
\[
f\partial_{\mu}f^{\nu\rho}\partial^{\mu}f_{\nu\rho}\,\,\,\,\,\,\,\,\,\,\,\,\,\,\text{vs}\,\,\,\,\,\,\,\,\,\,\,\,\,f^{\nu\rho}\partial^{\mu}f_{\nu\rho}\partial_{\mu}f.
\]
They have a very similar form and just differ by where the derivatives
act. As a result, they will give precisely the same contribution in
any diagram, this is a just a generic feature that follows from momentum
conservation and the Feynman rules. So no matter which higher amplitudes
we compute, there will always be at least one free parameter remaining.
This is expected: In general it is just not possible to uniquely determine
the Lagrangian, given the on-shell amplitudes only.

For the quartic vertex we have already highly constrained the set
of terms by the on-shell conditions. For higher amplitudes we would
need to add all possible terms containing traces and divergences.
At high enough order, they will contribute. Still, a similar phenomenon
as for the three-vertex will probably occur and we have to expect
that some parameters will remain free.

\section{Conclusion and outlook}

In this work we made an attempt to analyze the double copy from a
Lagrangian viewpoint, trying to make it more transparent. We set up
an effective Lagrangian based on some natural assumptions such that
its structure stays as close as possible to Yang-Mills theory. Then
we constrained it by requiring that the 3-point and 4-point amplitude
exactly match the prediction of the double copy. 

What we found was that this effective Lagrangian has a strong resemblance
with the teleparallel equivalent of Einstein gravity (TEGR). In fact,
up to fourth order we were able to reproduce perturbative TEGR up
to a few free real parameters. This is a strong indication that TEGR
based on torsion might be the better framework to understand the double
copy rather than ordinary Einstein gravity based on curvature. 

For the future, a few further directions remain to be explored:
\begin{itemize}
\item Of course the obvious extension is to include higher amplitudes as
well and see what kind of additional constraints emerge. This is in
principle straightforward although it requires the inclusion of a
large number of new terms and much higher computational effort.
\item Another direction is the analysis of gauge freedom within this context.
Here, we eliminated the freedom of having local Lorentz transformations
of the vielbein completely from the beginning and also did not impose
diffeomorphism invariance at the interacting level. This makes sense,
since in our ansatz we did not know that the resulting theory will
be gravity and hence tried to stay as general as possible. Still,
it could be interesting to see what kind of restrictions would arise
if some kind of gauge redundancy would be imposed.
\item This also naturally leads us to another point, namely making contact
with \cite{Borsten2021} and related work. Those references make heavy
use of homotopy algebras, so understanding TEGR from an $L_{\infty}$-perspective
may shed new light on the whole construction.
\item Lastly, a very interesting remaining problem is the extension of the
double copy from on-shell amplitudes to off-shell quantities. There
has been a lot of research concerning classical solutions, see \cite{Monteiro2014}
and references in \cite{Bern2024}. In constrast to Einstein gravity,
TEGR provides another off-shell quantity that can be compared, namely
a conserved, local energy-momentum-tensor. It would very instructive
to see if its structure can be related to the Yang-Mills energy momentum
tensor with some kind of double-copy procedure. 
\end{itemize}

\subsubsection*{Acknowledgements}

GS wants to thank Ashoke Sen for discussion and useful comments on
the draft. Moreover, the author acknowledges the use of ChatGPT for
setting up the symbolic calculations and proof reading. The work was
done under the support of the Department of Atomic Energy, Government
of India, project no. RTI4019. 

\appendix

\section{Spin projection operators and the graviton propagator}

Spin projection operators are a useful mathematical tool to decompose
and invert certain operators acting on tensor spaces. For instance,
the kinetic operator of any graviton theory is a symmetric linear
operator on the space of symmetric tensors in four dimensions. This
can be seen from the general schematic form of the action 
\begin{equation}
S\propto\int d^{4}x\,f^{\mu\nu}\mathcal{O}_{\mu\nu\rho\sigma}^{kin}f^{\rho\sigma}+\text{interaction terms}.
\end{equation}
$\mathcal{O}^{kin}$ can therefore be described as a ten-dimensional,
symmetric matrix, with possibly 55 independent entries in total. The
actual number of degrees of freedom is much lower though: Typically,
the kinetic term only involves derivatives and contractions, so its
Fourier transform is solely constructed out of $\eta_{\mu\nu}$ and
$p^{\mu}$. So we know that every Lorentz transformation which leaves
$p^{\mu}$ invariant, commutes with $\mathcal{O}^{kin}$. Hence by
Schur\textasciiacute s lemma, in every irreducible representation
of the little group of $p^{\mu}$, it has to be proportional to the
identity operator. Since it is an off-shell object, we can choose
$p^{\mu}$ to be timelike, such that the little group is just SO(3).
This tensor representation of SO(3) is decomposable into four irreducible
representations: one with spin two (5 degrees of freedom), one with
spin one (3 dof) and two scalars (1 dof each). It is now useful to
define the so-called \emph{spin projection operators }which\emph{
}project onto these four subspaces:\emph{ 
\begin{align}
 & P^{2}=\frac{1}{2}\left(\theta_{\mu\rho}\theta_{\nu\sigma}+\theta_{\mu\sigma}\theta_{\nu\rho}\right)-\frac{1}{3}\theta_{\mu\nu}\theta_{\rho\sigma}\nonumber \\
 & P^{1}=\frac{1}{2}\left(\theta_{\mu\rho}\omega_{\nu\sigma}+\theta_{\mu\sigma}\omega_{\nu\rho}+\theta_{\nu\rho}\omega_{\mu\sigma}+\theta_{\nu\sigma}\omega_{\mu\rho}\right)\nonumber \\
 & P_{s}^{0}=\frac{1}{3}\theta_{\mu\nu}\theta_{\rho\sigma}\nonumber \\
 & P_{w}^{0}=\omega_{\mu\nu}\omega_{\rho\sigma}
\end{align}
}with 
\begin{equation}
\theta_{\mu\nu}=\eta_{\mu\nu}-\frac{p_{\mu}p_{\nu}}{p^{2}}\,\,\,\mbox{and}\,\,\,\omega_{\mu\nu}=\frac{p_{\mu}p_{\nu}}{p^{2}}.
\end{equation}
It is easy to verify that they are all idempotent and satisfy 
\begin{equation}
P^{2}+P^{1}+P_{s}^{0}+P_{w}^{0}=1\,\,\,\mbox{and}\,\,\,P_{a}^{i}P_{b}^{j}=\delta^{ij}\delta_{ab}.
\end{equation}
$P^{2}$ is transverse and traceless, $P^{1}$ is longitudinal and
traceless, $P_{s}^{0}$ is the transverse trace part and $P_{w}^{0}$
is neither transverse nor traceless. Any kinetic operator has to be
proportional to unity in each subspace and must not mix subspaces
of different spin, however, it could potentially mix the two scalars.
We therefore additionally define the two intertwiners 
\begin{equation}
P_{sw}^{0}=\frac{1}{\sqrt{3}}\theta_{\mu\nu}\omega_{\rho\sigma}\,\,\,\,\mbox{and}\,\,\,\,P_{ws}^{0}=\frac{1}{\sqrt{3}}\omega_{\mu\nu}\theta_{\rho\sigma}.
\end{equation}
Now $\mathcal{O}^{kin}$ can be expanded as 
\begin{equation}
\mathcal{O}^{kin}=AP^{2}+BP^{1}+CP_{s}^{0}+DP_{w}^{0}+EP_{sw}^{0}+EP_{ws}^{0}\label{eq:general spin decompo}
\end{equation}
and we see that the 55 independent components got reduced to five
(for a symmetric operator, the coefficients of the two intertwiners
must coincide). In the case of Einstein gravity for instance we have
$A=p^{2},$ $C=-2p^{2}$ and all others vanishing, i. e. 
\begin{equation}
\mathcal{O}_{GR}^{kin}=p^{2}\left(P^{2}-2P_{s}^{0}\right).\label{eq:O_kin GR}
\end{equation}
Despite of simplification, the spin projectors have another advantage,
namely the possibility of inverting the kinetic term. For example,
$\mathcal{O}_{GR}^{kin}$ is per se not invertible, which is a simple
consequence of perturbative gravity having gauge redundancy. However,
we can invert (\ref{eq:O_kin GR}) on each invariant subspace, giving
rise to the \emph{reduced propagator}
\begin{equation}
\Pi_{red}=\frac{1}{p^{2}}\left(P^{2}-\frac{1}{2}P_{s}^{0}\right)=\frac{1}{2p^{2}}\left(\theta_{\mu\rho}\theta_{\nu\sigma}+\theta_{\mu\sigma}\theta_{\nu\rho}-\theta_{\mu\nu}\theta_{\rho\sigma}\right).\label{eq:GR propagator}
\end{equation}

To obtain the full propagator, we add the \emph{gauge fixing term}
\begin{equation}
\mathcal{L}_{gf}=-\frac{1}{2\xi}\left(\partial_{\mu}f^{\mu\nu}-\frac{1}{2}\partial^{\nu}f\right)\left(\partial_{\rho}f_{\,\,\,\nu}^{\rho}-\frac{1}{2}\partial_{\nu}f\right)
\end{equation}
to the Lagrangian. It enforces the \emph{De-Donder-gauge }condition
\begin{equation}
\left(\partial_{\mu}f^{\mu\nu}-\frac{1}{2}\partial^{\nu}f\right)=0
\end{equation}
while at the same time makes the kinetic operator invertible. The
resulting propagator for the choice $\xi=1$ is called \emph{Feynman
gauge propagator }and is given by 
\begin{equation}
\Pi_{GR}=\frac{1}{2p^{2}}\left(\eta_{\mu\rho}\eta_{\nu\sigma}+\eta_{\mu\sigma}\eta_{\nu\rho}-\eta_{\mu\nu}\eta_{\rho\sigma}\right).\label{eq:Feynman prop}
\end{equation}
If $\Pi_{GR}$ is sandwiched between two conserved sources, it can
be replaced by the reduced propagator,
\begin{equation}
\tau_{\mu\nu}\Pi_{GR}^{\mu\nu\rho\sigma}\tau_{\rho\sigma}=\tau_{\mu\nu}\Pi_{red}^{\mu\nu\rho\sigma}\tau_{\rho\sigma}.
\end{equation}
For the four point amplitude in the generic effective theory, computed
in section 5, we need to use the full propagator $\Pi_{GR}$ since
it is not guaranteed that it gets contracted with something conserved,
i.e., transverse.

\bibliographystyle{plain}
\bibliography{Teleparallel_double_copy}

@Article{Bern2024,
  author        = {Bern, Zvi and Carrasco, John Joseph and Chiodaroli, Marco and Johansson, Henrik and Roiban, Radu},
  journal       = {J. Phys. A},
  title         = {{The duality between color and kinematics and its applications}},
  year          = {2024},
  number        = {33},
  pages         = {333002},
  volume        = {57},
  archiveprefix = {arXiv},
  doi           = {10.1088/1751-8121/ad5fd0},
  eprint        = {1909.01358},
  primaryclass  = {hep-th},
  reportnumber  = {CERN-TH-2019-135, UCLA/TEP/2019/104, NUHEP-TH/19-11, UUITP-35/19, NORDITA 2019-079},
}

@Article{Bern2010,
  author        = {Bern, Zvi and Carrasco, John Joseph M. and Johansson, Henrik},
  journal       = {Phys. Rev. Lett.},
  title         = {{Perturbative Quantum Gravity as a Double Copy of Gauge Theory}},
  year          = {2010},
  pages         = {061602},
  volume        = {105},
  archiveprefix = {arXiv},
  doi           = {10.1103/PhysRevLett.105.061602},
  eprint        = {1004.0476},
  primaryclass  = {hep-th},
  reportnumber  = {UCLA-10-TEP-102, SACLAY-IPHT-T10-044},
}

@Article{Bern2010a,
  author        = {Bern, Zvi and Dennen, Tristan and Huang, Yu-tin and Kiermaier, Michael},
  journal       = {Phys. Rev. D},
  title         = {{Gravity as the Square of Gauge Theory}},
  year          = {2010},
  pages         = {065003},
  volume        = {82},
  archiveprefix = {arXiv},
  doi           = {10.1103/PhysRevD.82.065003},
  eprint        = {1004.0693},
  primaryclass  = {hep-th},
  reportnumber  = {UCLA-TEP-10-103, PUPT-2335},
}

@Article{Tolotti2013,
  author        = {Tolotti, Mathias and Weinzierl, Stefan},
  journal       = {JHEP},
  title         = {{Construction of an effective Yang-Mills Lagrangian with manifest BCJ duality}},
  year          = {2013},
  pages         = {111},
  volume        = {07},
  archiveprefix = {arXiv},
  doi           = {10.1007/JHEP07(2013)111},
  eprint        = {1306.2975},
  primaryclass  = {hep-th},
}

@Article{Brezina2026,
  author        = {Brezina, Sebastian and Boffo, Eugenia and Kr{\v{s}}{\v{s}}{\'a}k, Martin},
  journal       = {JHEP},
  title         = {{Teleparallel gravity from the principal bundle viewpoint}},
  year          = {2026},
  pages         = {182},
  volume        = {05},
  archiveprefix = {arXiv},
  doi           = {10.1007/JHEP05(2026)182},
  eprint        = {2507.23672},
  primaryclass  = {gr-qc},
}

@Article{Krssak2025,
  author        = {Kr{\v{s}}{\v{s}}{\'a}k, Martin},
  journal       = {Phys. Rev. Lett.},
  title         = {{Self-Excited Gravitational Instantons}},
  year          = {2025},
  number        = {20},
  pages         = {201501},
  volume        = {134},
  archiveprefix = {arXiv},
  doi           = {10.1103/PhysRevLett.134.201501},
  eprint        = {2408.01140},
  primaryclass  = {gr-qc},
}

@Article{Arcos2004,
  author        = {Arcos, H. I. and Pereira, J. G.},
  journal       = {Int. J. Mod. Phys. D},
  title         = {{Torsion gravity: A Reappraisal}},
  year          = {2004},
  pages         = {2193--2240},
  volume        = {13},
  archiveprefix = {arXiv},
  doi           = {10.1142/S0218271804006462},
  eprint        = {gr-qc/0501017},
}

@Book{Aldrovandi2013,
  author    = {Aldrovandi, Ruben and Pereira, Jos{\'e} Geraldo},
  publisher = {Springer},
  title     = {{Teleparallel Gravity}: {An Introduction}},
  year      = {2013},
  isbn      = {978-94-007-5142-2, 978-94-007-5143-9},
  doi       = {10.1007/978-94-007-5143-9},
}

@Article{LeDelliou2020,
  author        = {Le Delliou, M. and Huguet, E. and Fontanini, M.},
  journal       = {Phys. Rev. D},
  title         = {{Teleparallel theory as a gauge theory of translations: Remarks and issues}},
  year          = {2020},
  number        = {2},
  pages         = {024059},
  volume        = {101},
  archiveprefix = {arXiv},
  doi           = {10.1103/PhysRevD.101.024059},
  eprint        = {1910.08471},
  primaryclass  = {gr-qc},
}

@Article{Huguet2021,
  author        = {Huguet, E. and Le Delliou, M. and Fontanini, M. and Lin, Z. C.},
  journal       = {Phys. Rev. D},
  title         = {{Teleparallel gravity as a gauge theory: Coupling to matter using the Cartan connection}},
  year          = {2021},
  number        = {4},
  pages         = {044061},
  volume        = {103},
  archiveprefix = {arXiv},
  doi           = {10.1103/PhysRevD.103.044061},
  eprint        = {2008.13493},
  primaryclass  = {gr-qc},
}

@Article{Landry2023,
  author        = {Landry, Alexandre and Hoogen, Robert J. van den},
  journal       = {Universe},
  title         = {{Teleparallel Minkowski Spacetime with Perturbative Approach for Teleparallel Gravity on a Proper Frame}},
  year          = {2023},
  number        = {5},
  pages         = {232},
  volume        = {9},
  archiveprefix = {arXiv},
  doi           = {10.3390/universe9050232},
  eprint        = {2303.16089},
  primaryclass  = {gr-qc},
}

@Article{Kawai1986,
  author       = {Kawai, H. and Lewellen, D. C. and Tye, S. H. H.},
  journal      = {Nucl. Phys. B},
  title        = {{A Relation Between Tree Amplitudes of Closed and Open Strings}},
  year         = {1986},
  pages        = {1--23},
  volume       = {269},
  doi          = {10.1016/0550-3213(86)90362-7},
  reportnumber = {CLNS-85/667},
}

@Article{Bern2008,
  author        = {Bern, Z. and Carrasco, J. J. M. and Johansson, Henrik},
  journal       = {Phys. Rev. D},
  title         = {{New Relations for Gauge-Theory Amplitudes}},
  year          = {2008},
  pages         = {085011},
  volume        = {78},
  archiveprefix = {arXiv},
  doi           = {10.1103/PhysRevD.78.085011},
  eprint        = {0805.3993},
  primaryclass  = {hep-ph},
  reportnumber  = {UCLA-07-TEP-15},
}

@Article{Borsten2022,
  author        = {Borsten, Leron and Kim, Hyungrok and Jurco, Branislav and Macrelli, Tommaso and Saemann, Christian and Wolf, Martin},
  journal       = {PoS},
  title         = {{Colour-kinematics duality, double copy, and homotopy algebras}},
  year          = {2022},
  month         = {11},
  pages         = {426},
  volume        = {ICHEP2022},
  archiveprefix = {arXiv},
  doi           = {10.22323/1.414.0426},
  eprint        = {2211.16405},
  primaryclass  = {hep-th},
  reportnumber  = {DMUS-MP-22/22, EMPG-22-23},
}

@Article{Borsten2021,
  author        = {Borsten, Leron and Kim, Hyungrok and Jur{\v{c}}o, Branislav and Macrelli, Tommaso and Saemann, Christian and Wolf, Martin},
  journal       = {Fortsch. Phys.},
  title         = {{Double Copy from Homotopy Algebras}},
  year          = {2021},
  number        = {8-9},
  pages         = {2100075},
  volume        = {69},
  archiveprefix = {arXiv},
  doi           = {10.1002/prop.202100075},
  eprint        = {2102.11390},
  primaryclass  = {hep-th},
  reportnumber  = {DMUS-MP-21/04, EMPG-21-03},
}

@Article{Borsten2023,
  author        = {Borsten, Leron and Kim, Hyungrok and Jur{\v{c}}o, Branislav and Macrelli, Tommaso and Saemann, Christian and Wolf, Martin},
  journal       = {Nucl. Phys. B},
  title         = {{Tree-level color{\textendash}kinematics duality implies loop-level color{\textendash}kinematics duality up to counterterms}},
  year          = {2023},
  pages         = {116144},
  volume        = {989},
  archiveprefix = {arXiv},
  doi           = {10.1016/j.nuclphysb.2023.116144},
  eprint        = {2108.03030},
  primaryclass  = {hep-th},
  reportnumber  = {EMPG-21-12, DMUS-MP-21/12},
}

@Article{Borsten2023a,
  author        = {Borsten, Leron and Jurco, Branislav and Kim, Hyungrok and Macrelli, Tommaso and Saemann, Christian and Wolf, Martin},
  journal       = {Phys. Rev. D},
  title         = {{Tree-level color-kinematics duality from pure spinor actions}},
  year          = {2023},
  number        = {12},
  pages         = {126012},
  volume        = {108},
  archiveprefix = {arXiv},
  doi           = {10.1103/PhysRevD.108.126012},
  eprint        = {2303.13596},
  primaryclass  = {hep-th},
  reportnumber  = {DMUS-MP-23/05, EMPT-23-04},
}

@Article{Monteiro2014,
  author        = {Monteiro, Ricardo and O'Connell, Donal and White, Chris D.},
  journal       = {JHEP},
  title         = {{Black holes and the double copy}},
  year          = {2014},
  pages         = {056},
  volume        = {12},
  archiveprefix = {arXiv},
  doi           = {10.1007/JHEP12(2014)056},
  eprint        = {1410.0239},
  primaryclass  = {hep-th},
  reportnumber  = {EDINBURGH-2014-18},
}

@Article{Brezina2026a,
  author        = {Brezina, Sebastian and Kr{\v{s}}{\v{s}}{\'a}k, Martin},
  title         = {{Gauge aspects of teleparallel gravity and Noether's second theorem}},
  year          = {2026},
  month         = {11},
  archiveprefix = {arXiv},
  eprint        = {2611.17147},
  primaryclass  = {gr-qc},
}

\end{document}